\documentclass[american,twocolumn,prl]{revtex4-2}
\usepackage[T1]{fontenc}
\usepackage[latin9]{inputenc}
\usepackage{amsmath}
\usepackage{amssymb}
\usepackage{graphicx}
\PassOptionsToPackage{normalem}{ulem}
\usepackage{ulem}

\makeatletter

\providecommand{\tabularnewline}{\\}

\makeatother

\usepackage{babel}
\begin{document}
\title{Phase and amplitude evolution in the network of triadic interactions
of the Hasegawa-Wakatani system}
\author{Ö. D. Gürcan, J. Anderson, S. Moradi, A. Biancalani, P. Morel}
\affiliation{CNRS, LPP, Ecole Polytechnique}
\begin{abstract}
Hasegawa-Wakatani system, commonly used as a toy model of dissipative
drift waves in fusion devices is revisited with considerations of
phase and amplitude dynamics of its triadic interactions. It is observed
that a single resonant triad can saturate via three way phase locking
where the phase differences between dominant modes converge to constant
values as individual phases increase in time. This allows the system
to have approximately constant amplitude solutions. Non-resonant triads
show similar behavior only when one of its legs is a zonal wave number.
However when an additional triad, which is a reflection of the original
one with respect to the $y$ axis is included, the behavior of the
resulting triad pair is shown to be more complex. In particular, it
is found that triads involving small radial wave numbers (large scale
zonal flows) end up transferring their energy to the subdominant mode
which keeps growing exponentially, while those involving larger radial
wave numbers (small scale zonal flows) tend to find steady chaotic
or limit cycle states (or decay to zero). In order to study the dynamics
in a connected network of triads, a network formulation is considered
including a pump mode, and a number of zonal and non-zonal subdominant
modes as a dynamical system. It was observed that the zonal modes
become clearly dominant only when a large number of triads are connected.
When the zonal flow becomes dominant as a 'collective mean field',
individual interactions between modes become less important, which
is consistent with the inhomogeneous wave-kinetic picture. Finally,
the results of direct numerical simulation is discussed for the same
parameters and various forms of the order parameter are computed.
It is observed that nonlinear phase dynamics results in a flattening
of the large scale phase velocity as a function of scale in direct
numerical simulations.
\end{abstract}
\maketitle

\section{Introduction}

Two dimensional Hasegawa-Wakatani equations\citep{hasegawa:83} with
proper zonal response consist of an equation of plasma vorticity
\begin{equation}
\frac{\partial}{\partial t}\nabla^{2}\Phi+\hat{\mathbf{z}}\times\nabla\Phi\cdot\nabla\nabla^{2}\Phi=C\left(\widetilde{\Phi}-\widetilde{n}\right)+D_{\Phi}\left(\nabla^{2}\Phi\right)\label{eq:hw1}
\end{equation}
and an equation of continuity
\begin{equation}
\frac{\partial}{\partial t}n+\hat{\mathbf{z}}\times\nabla\Phi\cdot\nabla n+\kappa\partial_{y}\Phi=C\left(\widetilde{\Phi}-\widetilde{n}\right)+D_{n}\left(n\right)\;\text{,}\label{eq:hw2}
\end{equation}
with the $E\times B$ velocity defined as $\boldsymbol{v}_{E}=\hat{\mathbf{z}}\times\nabla\Phi$
in normalized form and $\widetilde{\Phi}=\Phi-\left\langle \Phi\right\rangle $
where $\left\langle \Phi\right\rangle $ denotes averaging in $y$
(i.e. poloidal) direction. Here $n$ is the fluctuating particle density
normalized to a background density $n_{0}$, $\Phi$ is the electrostatic
potential normalized to $T/e$, $\kappa$ is the diamagnetic velocity
normalized to speed of sound, $C$ is the so called adiabaticity parameter,
which is a measure of the electron mobility and $D_{\Phi}$ and $D_{n}$
are dissipation functions for vorticity and particle density respectively.
For fluctuations we have $D_{\Phi}\left(\nabla^{2}\widetilde{\Phi}\right)=\nu\nabla^{4}\widetilde{\Phi}$
from kinematic viscosity, whereas for the zonal flows $D_{\Phi}\left(\nabla^{2}\overline{\Phi}\right)=-\nu_{ZF}\nabla^{2}\overline{\Phi}$
from large scale friction. Unless the system represents a renormalized
formulation, $D_{n}$ should actually be zero, however here we include
it for completeness and numerical convenience and take it to have
the same form as the vorticity dissipation with diffusion $D_{n}\left(\widetilde{n}\right)=D\nabla^{2}\widetilde{n}$
and particle loss $D\left(\overline{n}\right)=-D_{ZF}\overline{n}$.

The Hasegawa Wakani model, was initially devised as a simple, nonlinear
model of dissipative drift wave turbulence in tokamak plasmas. It
has the same nonlinear structure as the passive scalar turbulence
\citep{lesieur:85} -with vorticity evolving according to 2D Navier-Stokes
equations- or more complex problems such as rotating convection \citep{busse:80,currie:16}.
From a plasma physics perspective it can be considered as the minimum
non-trivial model for plasma turbulence, since it has i) linear instability
(e.g. Hasegawa-Mima model does not \citep{hasegawa:78}), ii) finite
frequency (so that resonant interactions are possible \citep{connaughton:15}),
and iii) a proper treatment of zonal flows\citep{diamond:05}. The
model is well known to generate high levels of large scale zonal flows,
especially for $C\gtrsim1$ \citep{holland:07,pushkarev:13,zhang:20}.
It has been studied in detail for many problems in fusion plasmas
including dissipative drift waves in tokamak edge \citep{scott:88,koniges:92},
subcritical turbulence\citep{friedman:15}, trapped ion modes \citep{sarto:17},
intermittency \citep{bos:10,anderson:17}, closures \citep{gang:91b,hu:97,singh:21},
feedback control \citep{goumiri:13}, information geometry \citep{anderson:20}
and machine learning \citep{heinonen:20}. Variations of the Hasegawa-Wakatani
model are regularly used for describing turbulence in basic plasma
devices\citep{kasuya:07,vaezi:17,donnel:18}.

Formation of large scale structures, in particular Zonal flows in
drift wave turbulence is one of the key issues in the study of turbulence
in fusion plasmas, which can be formulated in terms of modulational
instability of either a gas of drift wave turbulence using the wave
kinetic formulation \citep{smolyakov:1999} or a small number of drift
modes\citep{chen:00}, resulting in various forms of complex amplitude
equations such as the celebrated nonlinear Schrödinger equation (NLS)
\citep{champeaux:01}. It is also common to talk about zonal flows
as resulting from a process of inverse cascade\citep{manz:09,stroth:11}and
their back reaction on turbulence\citep{biglari:90,terry:00} results
in predator-prey dynamics, possibly leading up to the low to high
confinement transition in tokamaks\citep{malkov:01,kim:03b}. While
the role of the complex phases in nonlinear evolution of the amplitudes,
especially in the context of structure formation, for example as in
the case of soliton formation in NLS, was always well known, its particular
importance for zonal flow formation in toroidal geometry has been
underlined recently\citep{guo:16}.

Here we revisit the Hasegawa-Wakatani system, with proper zonal response,
as a minimum system that allows a description of zonal flow formation
in drift wave turbulence, and study interactions between various number
of modes from three wave interactions to the full spectrum of modes
described by direct numerical simulations, focusing in particular
on phase dynamics and the possibility of phase locking and synchronization.
It turns out that while resonant three wave interactions involving
unstable and damped modes favor phase locking (i.e. a state where
the differences between individual phases remain roughly constant
as they increase together), interactions involving zonal flows (i.e.
four wave interactions including the triad reflected with respect
to the $y$ axis), seems to have a complicated set of possible outcomes
depending on if the zonal flow wave number is larger or smaller than
the pump wave-number. It therefore becomes critical to study a ``network''
of connected triads in order to see the collective effects of a number
of triads on the evolution of zonal flows and of the relative phases
between modes. Two different network configurations are considered:
that with a single $k_{y}$ and many different $q$'s, and that with
a single $q$ but many different $k_{y}$'s. Note that the algorithm
that we use computes all possible interactions between the modes in
a given collection of triads and then computes the interaction coefficients
and evolves the system nonlinearly according to those.

Finally we consider the results from direct numerical simulations
(DNS) using a pseudo-spectral 2D Hasegawa-Wakatani solver. The DNS
and the network models correspond exactly, in the sense that if we
consider an $N_{x}\times N_{y}$ grid and consider all the possible
triads in such a grid and solve this problem using our network solver,
we obtain exactly the same problem (including the boundary conditions
that are periodic) as the DNS. The results of the DNS show qualitatively
similar behavior to the two network models that we considered. However
looking at the evolution of phases, we observe a nonlinear flattening
of the phase velocity for large scales computed as a function of $x$,
suggesting nonlinear structure formation in the classical sense of
nonlinearity balancing dispersion resulting in a constant velocity
propagation at least for large r scale structures. These vortex-like
structures that move at a constant velocity are also clearly visible
in the time evolution of density and vorticity fields.

The rest of the paper is organized as follows. In the remainder of
the introduction, the Hasegawa-Wakatani system is reformulated in
terms of its linear eigenmodes and the amplitude and phase equations
for these eigenmodes, writing out explicitly the nonlinear terms that
appear in this formulation. In Section II, different types of interactions
among dissipative drift waves are considered using these linear eigenmodes,
starting with the basic three wave interaction. After showing that
there is no qualitative difference between a near resonant and an
exactly resonant (within numerical accuracy) triad, the details of
the phase dynamics of such a single triad are discussed. In Section
III, the interaction with zonal flows are considered. It is noted
that when we consider a triad and its reflection with respect to its
pump wave-number together as a pair, the behavior of the system is
qualitatively different from the single triad case. After a discussion
of order parameters for this system, a network formulation is considered
and the results from such a network model is presented. Finally the
reslts from direct numerical simulations of the Hasegawa-Wakatani
system is discussed and compared with those earlier results based
on reduced number of triads. Section IV is conclusion.

\subsection{Linear Eigenmodes}

We can write the Hasegawa-Wakatani system in Fourier space for non-zonal
modes (i.e. $k_{y}\neq0$) as follows:
\begin{equation}
\partial_{t}\Phi_{k}+\left(A_{k}-B_{k}\right)\Phi_{k}=\frac{C}{k^{2}}n_{k}+N_{\Phi k}\label{eq:hwft1}
\end{equation}
\begin{equation}
\partial_{t}n_{k}+\left(A_{k}+B_{k}\right)n_{k}=\left(C-i\kappa k_{y}\right)\Phi_{k}+N_{nk}\label{eq:hwft2}
\end{equation}
where 
\begin{equation}
A_{k}=\frac{1}{2}\left[\left(Dk^{2}+C\right)+\left(\frac{C}{k^{2}}+\nu k^{2}\right)\right]\label{eq:A}
\end{equation}
and
\begin{equation}
B_{k}=\frac{1}{2}\left[\left(Dk^{2}+C\right)-\left(\frac{C}{k^{2}}+\nu k^{2}\right)\right]\;\text{.}\label{eq:B}
\end{equation}
using the simpler notation $\Phi_{\mathbf{k}}\rightarrow\Phi_{k}$
and defining:

\begin{equation}
N_{nk}=\frac{1}{2}\sum_{\triangle}\hat{\mathbf{z}}\times\mathbf{p}\cdot\mathbf{q}\left(\Phi_{p}^{*}n_{q}^{*}-\Phi_{q}^{*}n_{p}^{*}\right)\label{eq:Nnk}
\end{equation}
 and 
\begin{equation}
N_{\Phi k}=\frac{1}{2}\sum_{\triangle}\frac{\hat{\mathbf{z}}\times\mathbf{p}\cdot\mathbf{q}\left(q^{2}-p^{2}\right)\Phi_{p}^{*}\Phi_{q}^{*}}{k^{2}}\;\text{.}\label{eq:Nphik}
\end{equation}

Diagonalizing the linear terms we can write:
\begin{equation}
\partial_{t}\xi_{k}^{\pm}+i\omega_{k}^{\pm}\xi_{k}^{\pm}=N_{\xi k}^{\pm}\label{eq:diag}
\end{equation}
with the complex eigen-frequencies $\omega_{k}^{\pm}=\omega_{rk}^{\pm}+i\gamma_{k}^{\pm}$
that can be written as:
\[
\omega_{k}^{\pm}=\Omega_{k}^{\pm}-iA_{k}
\]
with 
\begin{equation}
\Omega_{k}^{\pm}=\pm\left(\sigma_{k}\sqrt{\frac{H_{k}-G_{k}}{2}}+i\sqrt{\frac{H_{k}+G_{k}}{2}}\right)\label{eq:Omk}
\end{equation}
where $\sigma_{k}=\text{sign}\left(\kappa k_{y}\right)$,
\begin{equation}
H_{k}=\sqrt{G_{k}^{2}+C^{2}\kappa^{2}k_{y}^{2}/k^{4}}\;\text{,}\label{eq:H}
\end{equation}
and
\begin{equation}
G_{k}\equiv\left(B_{k}^{2}+\frac{C^{2}}{k^{2}}\right)\;\text{.}\label{eq:G}
\end{equation}
This allows us to write the two linear eigenmodes as:
\begin{equation}
\xi_{k}^{s_{k}}=n_{k}+\frac{k^{2}}{C}\left[B_{k}-i\Omega_{k}^{s_{k}}\right]\Phi_{k}\;\text{.}\label{eq:xikp}
\end{equation}
where $s_{k}=\pm$. The nonlinear terms in (\ref{eq:diag}) become:
\begin{equation}
N_{\xi k}^{s_{k}}=N_{nk}+\frac{k^{2}}{C}\left(B_{k}-i\Omega_{k}^{s_{k}}\right)N_{\Phi k}\;\text{,}\label{eq:Nxik}
\end{equation}
and the inverse transforms can be written as:
\begin{equation}
\Phi_{k}=\frac{i}{2}\frac{C}{k^{2}}\sum_{s_{k}}\frac{\xi_{k}^{s_{k}}}{\Omega_{k}^{s_{k}}}\label{eq:inv1}
\end{equation}
\begin{equation}
n_{k}=-\frac{i}{2}\sum_{s_{k}}\frac{1}{\Omega_{k}^{s_{k}}}\left(B_{k}+i\Omega_{k}^{s_{k}}\right)\xi_{k}^{s_{k}}\;\text{.}\label{eq:inv2}
\end{equation}
Considering the inviscid limit, $\left\{ D,\nu\right\} \rightarrow0$
and $k_{y}\rightarrow O\left(\epsilon\right)$, where we keep terms
only up to $O\left(\epsilon\right)$ we obtain:$k$
\[
\xi_{k}^{+}=n_{k}+\left(k^{2}-i\frac{\kappa k_{y}}{2A_{k}^{2}}\right)\Phi_{k}
\]
\[
\xi_{k}^{-}=n_{k}-\left(1+i\frac{\kappa k_{y}}{2A_{k}^{2}}\right)\Phi_{k}\;\text{,}
\]
which means that one could loosely refer to these two modes as the
potential vorticity mode (i.e. $\xi_{k}^{+}=n_{k}+k^{2}\Phi_{k}$)
and the non-adiabatic electron density mode (i.e. $\xi_{k}^{-}=n_{k}-\Phi_{k}$),
somewhat similar to the real space decomposition used in Ref. \citealp{stoltzfus-dueck:13}.
Since the equations are already diagonal for $k_{y}=0$ modes, we
can use $\xi_{k}^{+}=k^{2}\Phi_{k}$ and $\xi_{k}^{-}=n_{k}$ for
those (or $\overline{\Phi}_{k}$ and $\overline{n}_{k}$ explicitly
as we will do below).

Notice that the two eigenmodes in (\ref{eq:xikp}) are not orthogonal.
They have the same frequencies (in opposite directions) but different
growth rates with $\gamma_{k}^{+}>\gamma_{k}^{-}$ (with $\gamma_{k}^{-}<0$,
while $\gamma_{k}^{+}$ can be positive or negative depending on the
wave-number). The full nonlinear initial value problem can be solved
using linear eigenmodes by first computing $\xi_{k}^{s_{k}}\left(0\right)$
from (\ref{eq:xikp}), and then advancing those to $\xi_{k}^{s_{k}}\left(t\right)$
using (\ref{eq:diag}), where the linear matrix is now diagonal (but
the nonlinear coupling terms are rather complicated), and finally
going back to compute $\Phi_{k}\left(t\right)$ and $n_{k}\left(t\right)$
using (\ref{eq:inv1}-\ref{eq:inv2}). Obviously, this approach does
not involve any kind of approximation.

\subsection{Amplitude and Phase Equations\label{subsec:Amplitude-and-Phase}}

Substituting $\xi_{k}^{\pm}=\chi_{k}^{\pm}e^{i\phi_{k}^{\pm}}$ into
(\ref{eq:diag}), we get:
\begin{equation}
\partial_{t}\left(\chi_{k}^{\pm}e^{i\phi_{k}^{\pm}}\right)+i\omega_{k}^{\pm}\chi_{k}^{\pm}e^{i\phi_{k}^{\pm}}=\left|N_{\xi k}^{\pm}\right|e^{i\phi_{k}^{N_{\xi\pm}}}\;\text{,}\label{eq:phamp}
\end{equation}
taking the real part we obtain the amplitude equations:
\begin{equation}
\left(\partial_{t}-\gamma_{k}^{\pm}\right)\chi_{k}^{\pm}=\left|N_{\xi k}^{\pm}\right|\cos\left(\phi_{k}^{N_{\xi\pm}}-\phi_{k}^{\pm}\right)\label{eq:amp}
\end{equation}
and taking the imaginary part and dividing by $\chi_{k}^{\pm}$ we
get the phase equations:
\begin{equation}
\partial_{t}\phi_{k}^{\pm}=-\omega_{kr}^{\pm}+\frac{\left|N_{\xi k}^{\pm}\right|}{\chi_{k}^{\pm}}\sin\left(\phi_{k}^{N_{\xi\pm}}-\phi_{k}^{\pm}\right)\;\text{.}\label{eq:ph}
\end{equation}
The form of the amplitude equation (\ref{eq:amp}) means that the
fixed point for the amplitude evolution is determined by the phase
difference between $N_{\xi k}^{s_{k}}$ and $\xi_{k}^{s_{k}}$ for
each $s_{k}$. However such a fixed point keeps evolving since the
phases themselves increase linearly with the linear frequency while
being deformed by the nonlinear terms. Note that if the nonlinear
phase is dominated by a slowly evolving mean phase (could be the case
if the nonlinear interactions are dominated by the theractions with
a zonal flow), the individual phases will be attracted to this nonlinear
mean phase, since if the individual phase is behind the nonlinear
phase, the $\sin\left(\phi_{k}^{N_{\xi\pm}}-\phi_{k}^{\pm}\right)$
will be positive, causing the individual phase to accelerate, whereas
if the individual phase is ahead of the nonlinear phase it will be
slowed down. However since we have linear frequencies it is impossible
for individual phases to become phase locked directly with the slow
nonlinear phase. Instead the nonlinear term plays a role akin to that
of the ponderomotive force in parametric instability.

\subsection{Nonlinear Terms}

In order to compute $N_{\xi k}^{\pm}$ in terms of $\xi_{k}^{\pm}$,
we need to go back to $\Phi_{k}$ and $n_{k}$ using (\ref{eq:inv1}-\ref{eq:inv2}),
compute the nonlinear terms (\ref{eq:Nnk}-\ref{eq:Nphik}) using
those and combine them as in (\ref{eq:Nxik}). They can then be written
in the form:
\begin{figure}
\begin{centering}
\includegraphics[width=1\columnwidth]{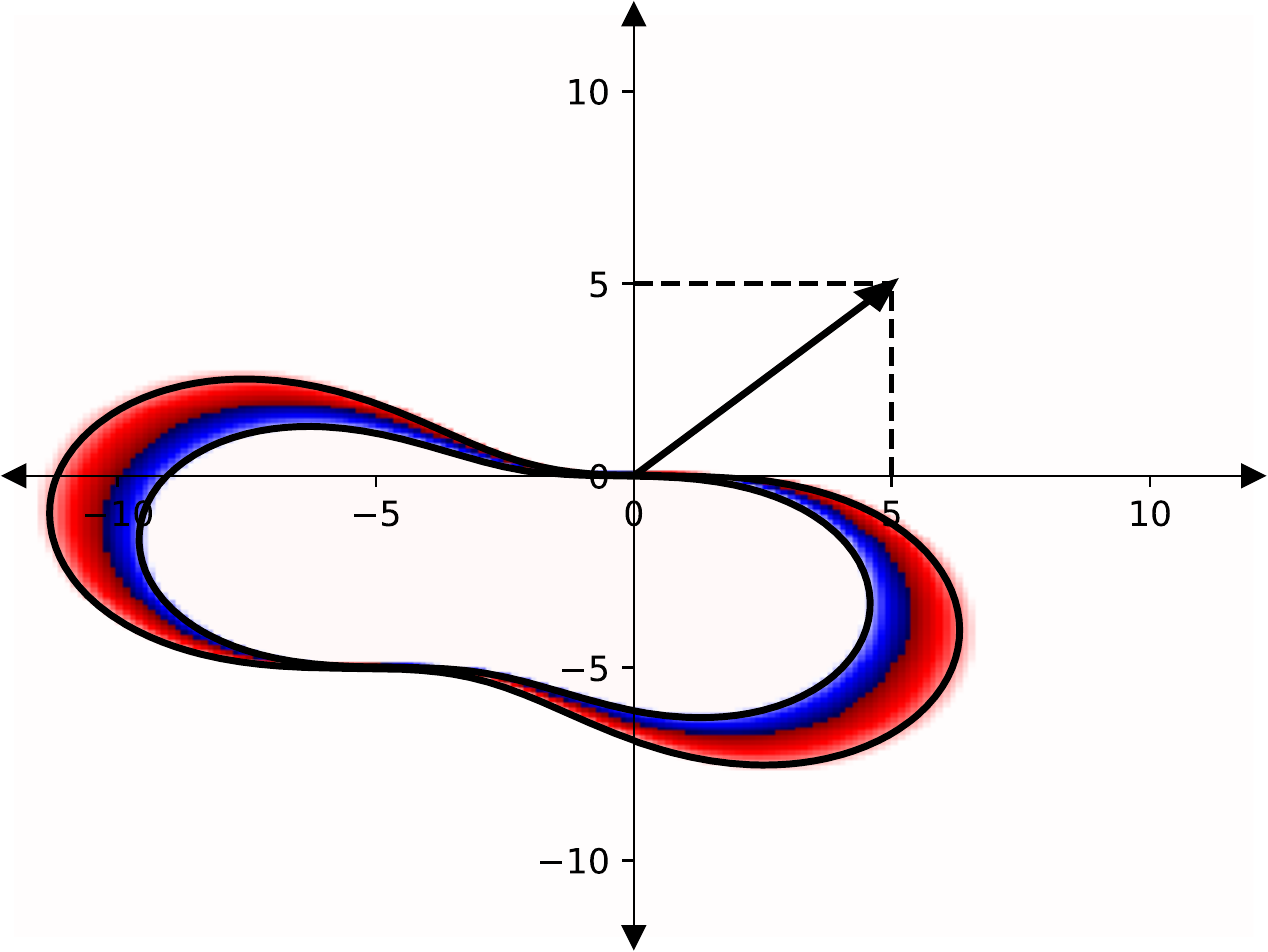}
\par\end{centering}
\caption{\label{fig:resman1}The resonance manifold $\Delta\omega=\omega_{\mathbf{k}r}^{+}+\omega_{\mathbf{p}r}^{+}+\omega_{\mathbf{q}r}^{+}=0$
of the Hasegawa-Wakatani system for the case $C=1.0$, $\kappa=0.2$,
$\nu=D=10^{-3}$ is shown corresponding to the wave vector $\mathbf{k}=\left(5,5\right)$
that is shown explicitly. Any $\mathbf{p}$ that falls onto the region
inside the resonance manifold (shown here with a finite width of $\pm0.04$
with $\Delta\omega>0$ in red and $\Delta\omega<0$ in blue if in
color) gives $\Delta\omega\approx0$ (with $\mathbf{q}=-\mathbf{k}-\mathbf{p}$).
As discussed in the text, because of the fact that the $\left(+\right)$
and $\left(-\right)$ modes have the same frequency (but opposite
direction of propagation in $y$ direction) all possible combinations
of $\left(+\right)$ and $\left(-\right)$ modes resonate on the same
manifold.}
\end{figure}
\begin{equation}
N_{\xi k}^{s_{k}}=\frac{1}{2}\sum_{\triangle}\sum_{s_{p},s_{q}}M_{kpq}^{s_{k}s_{p}s_{q}}\xi_{p}^{s_{p}*}\xi_{q}^{s_{p}*}\label{eq:nl}
\end{equation}
in terms of the linear eigenmodes, where the sum is over $s_{p},s_{q}=\left\{ \left(+,+\right),\left(+,-\right),\left(-,+\right),\left(-,-\right)\right\} $
for $s_{k}=\left(+,-\right)$. The nonlinear interaction coefficients
in (\ref{eq:nl}) can be written (i.e. between 3 non-zonal modes)
as:

\begin{align}
M_{\xi kpq}^{s_{k}s_{p}s_{q}} & =m_{kpq}^{s_{k}s_{p}s_{q}}\bigg[q^{2}\left(B_{q}-i\Omega_{q}^{s_{q}*}\right)\nonumber \\
 & -p^{2}\left(B_{p}-i\Omega_{p}^{s_{p}*}\right)-\left(q^{2}-p^{2}\right)\left(B_{k}-i\Omega_{k}^{s_{k}}\right)\bigg]\label{NLT}
\end{align}
where
\[
m_{kpq}^{s_{k}s_{p}s_{q}}\equiv\frac{C\hat{\mathbf{z}}\times\mathbf{p}\cdot\mathbf{q}}{4\Omega_{p}^{s_{p}*}\Omega_{q}^{s_{q}*}q^{2}p^{2}}
\]
and $\Omega_{k}^{\pm}$ is given in (\ref{eq:Omk}).

Note that these coefficients are complex, and have different phases
in general. In other words the explicit forms of (\ref{eq:ph}) can
be written as:
\begin{align}
\partial_{t}\phi_{k}^{s_{k}} & =-\omega_{kr}^{s_{k}}+\sum_{\triangle}\sum_{s_{p},s_{q}}\frac{\left|M_{\xi kpq}^{s_{k}s_{p}s_{q}}\right|\left|\xi_{p}^{s_{p}}\right|\left|\xi_{q}^{s_{q}}\right|}{\left|\xi_{k}^{s_{k}}\right|}\nonumber \\
 & \times\sin\left(\theta_{M_{kpq}}^{s_{k}s_{p}s_{q}}-\phi_{p}^{s_{p}}-\phi_{q}^{s_{q}}-\phi_{k}^{s_{k}}\right)\label{eq:phase}
\end{align}
where $\theta_{M_{kpq}}^{s_{k}s_{p}s_{q}}$ is the phase of the nonlinear
interaction coefficient $M_{\xi kpq}^{s_{k}s_{p}s_{q}}$.

\begin{figure}
\begin{centering}
\includegraphics[width=1\columnwidth]{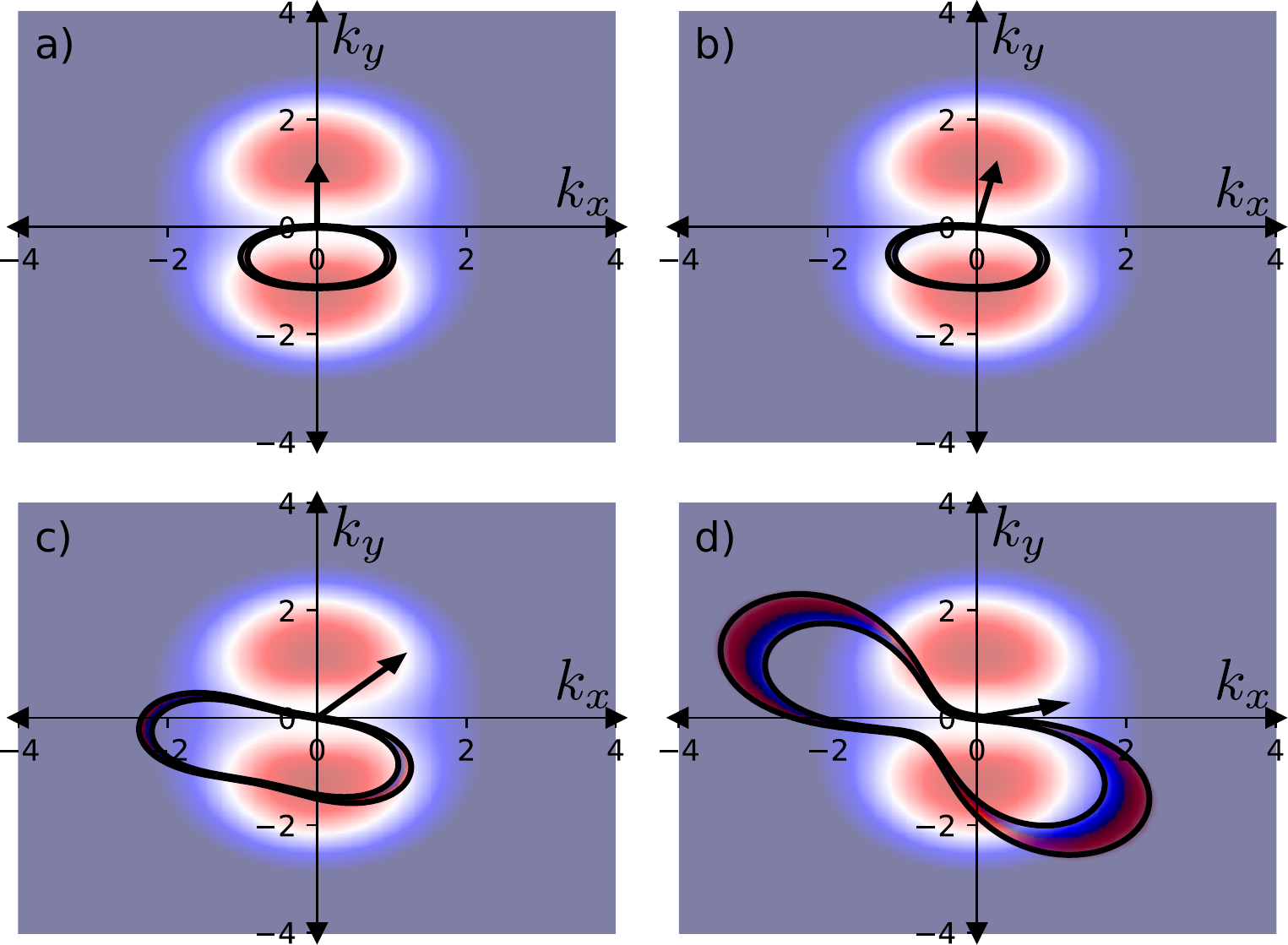}
\par\end{centering}
\caption{\label{fig:resman2}The resonance manifold, shown on top of the growth
rate where red corresponds to $\gamma_{\mathbf{k}}^{+}>0$ and blue
to $\gamma_{\mathbf{k}}^{+}<0$ for a) the most unstable model on
the grid $\mathbf{k}=\left(0,1.125\right)$, b) a nearby mode with
a small $k_{x}$ component $\mathbf{k}=\left(0.250,1.125\right)$,
c) a mode with $k_{x}=k_{y}$ that is $\mathbf{k}=\left(1.125,1.125\right)$
and finally d) a mode that has $k_{x}\gg k_{y}$ with $\mathbf{k}=\left(1.125,0.125\right)$.}
\end{figure}

\section{Interactions among drift waves}

\subsection{Three wave interactions}

Consider three separate modes $k$, $p$ and $q$ that satisfy the
triadic interaction condition $\mathbf{k}+\mathbf{p}+\mathbf{q}=0$,
possibly in the presence of other modes. The nonlinear term for the
wave number $k$ can then be divided into the interaction with the
pair $p$ and $q$, and the interaction with the rest of the modes
(if they exist). If the three wave interaction that we consider is
resonant, slightly off-resonance, or completely non-resonant, its
evolution is likely to be different, which can be considered as different
scenarios. It may also be possible to model the effects of rest of
the modes as background forcing, modification of the linear terms
(\emph{à la} eddy damping) or simply as stochastic noise. Thus separating
the nonlinear term into the interaction with the pair $p$ and $q$
(i.e. $N_{\xi kpq}^{\pm}$) and the interaction with the rest of the
modes (i.e. $\delta N_{\xi kpq}^{\pm}$), we can write:
\begin{equation}
\partial_{t}\xi_{k}^{\pm}+i\omega_{k}^{\pm}\xi_{k}^{\pm}=N_{\xi kpq}^{\pm}+\delta N_{\xi kpq}^{\pm}\label{eq:xik}
\end{equation}
where
\begin{align*}
N_{\xi kpq}^{\pm}= & M_{\xi kpq}^{\pm++}\xi_{p}^{+*}\xi_{q}^{+*}+M_{\xi kpq}^{\pm+-}\xi_{p}^{+*}\xi_{q}^{-*}\\
 & +M_{\xi kpq}^{\pm-+}\xi_{p}^{-*}\xi_{q}^{+*}+M_{\xi kpq}^{\pm--}\xi_{p}^{-*}\xi_{q}^{-*}
\end{align*}
with $M_{\xi kpq}^{\pm\pm\pm}$ being (complex) nonlinear interaction
coefficients, and
\[
\delta N_{\xi kpq}^{\pm}=N_{\xi k}^{\pm}-N_{\xi kpq}^{\pm}\;\text{.}
\]
The $p$ and $q$ modes evolve similarly:
\begin{equation}
\partial_{t}\xi_{p}^{\pm}+i\omega_{p}^{\pm}\xi_{p}^{\pm}=N_{\xi pqk}^{\pm}+\delta N_{\xi pqk}^{\pm}\label{eq:xip}
\end{equation}
\begin{equation}
\partial_{t}\xi_{q}^{\pm}+i\omega_{q}^{\pm}\xi_{q}^{\pm}=N_{\xi qkp}^{\pm}+\delta N_{\xi qkp}^{\pm}\label{eq:xiq}
\end{equation}
Notice that, since there are two eigenmodes (\ref{eq:xik}-\ref{eq:xiq})
represent 6 equations. One can therefore consider resonances between
$3$ growing modes, $2$ growing modes and a damped mode, or a growing
mode and $2$ damped modes etc. However since the frequencies are
the same with opposing signs, and due to the condition that the flow
field is real, we have both $k_{y}$ and $-k_{y}$ components, whenever
we have a resonance say of the form $\omega_{\mathbf{k}}^{+}+\omega_{\mathbf{p}}^{+}+\omega_{\mathbf{q}}^{+}=0$,
(with $\mathbf{k}+\mathbf{p}+\mathbf{q}=0$), we also have $\omega_{\mathbf{k}}^{-}+\omega_{\mathbf{p}}^{-}+\omega_{\mathbf{q}}^{-}=0$,
$\omega_{\mathbf{k}}^{+}-\omega_{-\mathbf{p}}^{-}-\omega_{-\mathbf{q}}^{-}=0$
or $\omega_{-\mathbf{k}}^{-}-\omega_{\mathbf{p}}^{+}-\omega_{\mathbf{q}}^{+}=0$
etc. In other words, whenever we have a resonance for three $+$ modes,
we also have all the other combinations. The form of the resonance
manifold can be seen in figures \ref{fig:resman1} and \ref{fig:resman2},
for $C=1$, $\kappa=0.2$, and $\nu=D=10^{-3}$, which we will refer
to as the ``$C=1$ case''.

The three wave interaction system (\ref{eq:xik}-\ref{eq:xiq}) can
be implemented numerically without much difficulty by dropping the
$\delta N_{\xi}$ terms above. One can also formulate the same three
wave interaction problem in the original variables $\Phi_{k}$, $\Phi_{p}$,
$\Phi_{q}$, $n_{k}$, $n_{p}$ and $n_{q}$ using the form (\ref{eq:hwft1}-\ref{eq:hwft2})
before the transformation, and then transform the result using (\ref{eq:xikp}).
Obviously those two approaches are numerically equivalent and naturally
they give exactly the same results. We used this to verify that the
eigenmode computation was correct. While in general it is unclear
if the eigenmode formulation provides any concrete advantage apart
from diagonalizing the linear system, the advantage becomes self-evident
if the resulting fluctuations have $\xi_{k}^{+}\gg\xi_{k}^{-}$ and
we can drop the $\xi_{k}^{-}$ mode for example.
\begin{figure}
\begin{centering}
\includegraphics[width=1\columnwidth]{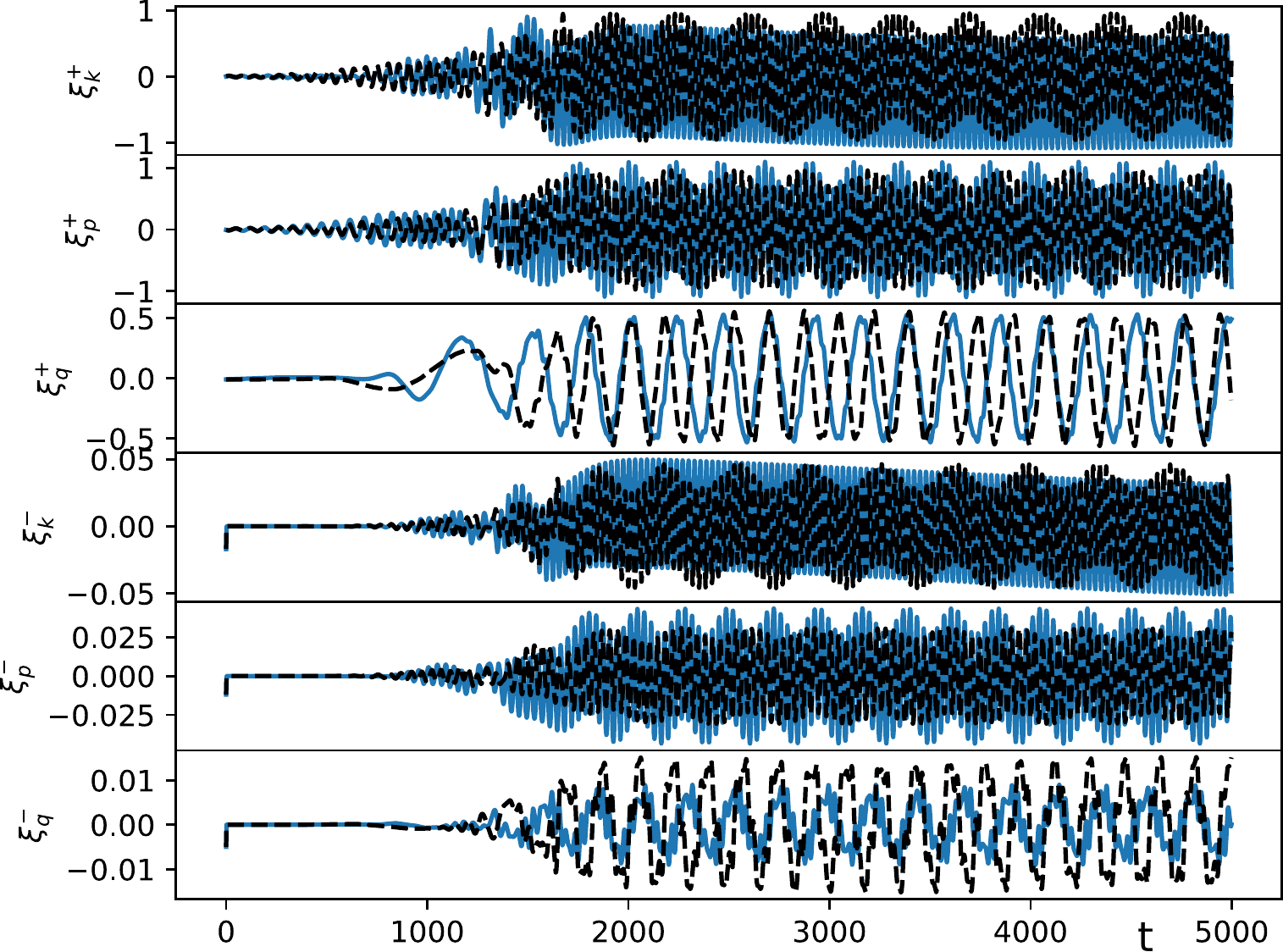}
\par\end{centering}
\caption{\label{fig:comparison}Comparison between exact or near resonances,
with real parts of each eigenmode shown for each wave number as labeled
on the left side of the figure. The solid line is the exact (i.e.
$\Delta\omega\approx2\times10^{-15}$) resonance of $\mathbf{k}=\left(0,1.125\right)$
with $\mathbf{p}=\left(-0.5,-1.0632325265492\right)$ whereas the
dashed line is the near resonance with $\mathbf{p}=\left(-0.5,-1.0\right)$
and $\Delta\omega\approx0.01$. While some details change, the overall
behavior, and saturation levels are actually very similar.}
\end{figure}

\subsubsection{Is there a difference between exact and near resonances?}

We first pick a primary wave-number $\mathbf{k}=\left(0,1.125\right)$
which is the linearly most unstable mode on a grid with $dk_{x}=dk_{y}=0.125$
for the $C=1.0$ case and consider the resonance manifold as shown
in figure \ref{fig:resman2}a in order to pick a second wave-number
$\mathbf{p}=\left(-0.5,-1.0\right)$ as the point on the $k$-space
grid that is closest to the resonance manifold. The third wave-number
$\mathbf{q}$ is computed from $\mathbf{k}+\mathbf{p}+\mathbf{q}=0$.
While a direct numerical simulation only has the wave-numbers on grid
points, a three wave equation solver is not constrained to such a
grid. We can instead compute $\mathbf{p}$ to be exactly on the resonance
manifold -at least within some numerical precision- for example by
choosing $\mathbf{p}=\left(-0.5,-1.0632325265492\right)$. Solving
the three wave equations numerically, using these slightly different
sets of wave-numbers, we find that having exact resonance or near
resonance (i.e. $\Delta\omega\approx2\times10^{-15}$ vs. $\Delta\omega\approx0.01$
) does not make much difference in terms of time evolution (see figure
), while picking something like $\mathbf{p}=\left(-0.5,-1.5\right)$,
which gives $\Delta\omega\approx0.07$ (with $\omega_{k}\approx0.1$
for comparison) gives a completely different time evolution, where
one of the modes keeps growing linearly without being able to couple
to the other two. We verified this for a bunch of different sets of
wave numbers, and while there are some differences in detail, generally
both exactly resonant or near resonant triads lead to saturation but
non-resonant triads can not saturate, possibly due to lack of efficient
interactions. The boundary between what can be considered a near resonant
vs. non-resonant interaction can actually be defined using this criterion.
In particular, it seems that the triads with one of the frequencies
much smaller than the other two (i.e. $\omega_{q}\ll\omega_{p}\sim\omega_{k}$)
tend to support larger overall $\Delta\omega$, and nonetheless reach
saturation. However it is not clear whether these observations from
a single triad continue to hold when many triads are interacting with
each other.
\begin{figure}
\begin{centering}
\includegraphics[width=1\columnwidth]{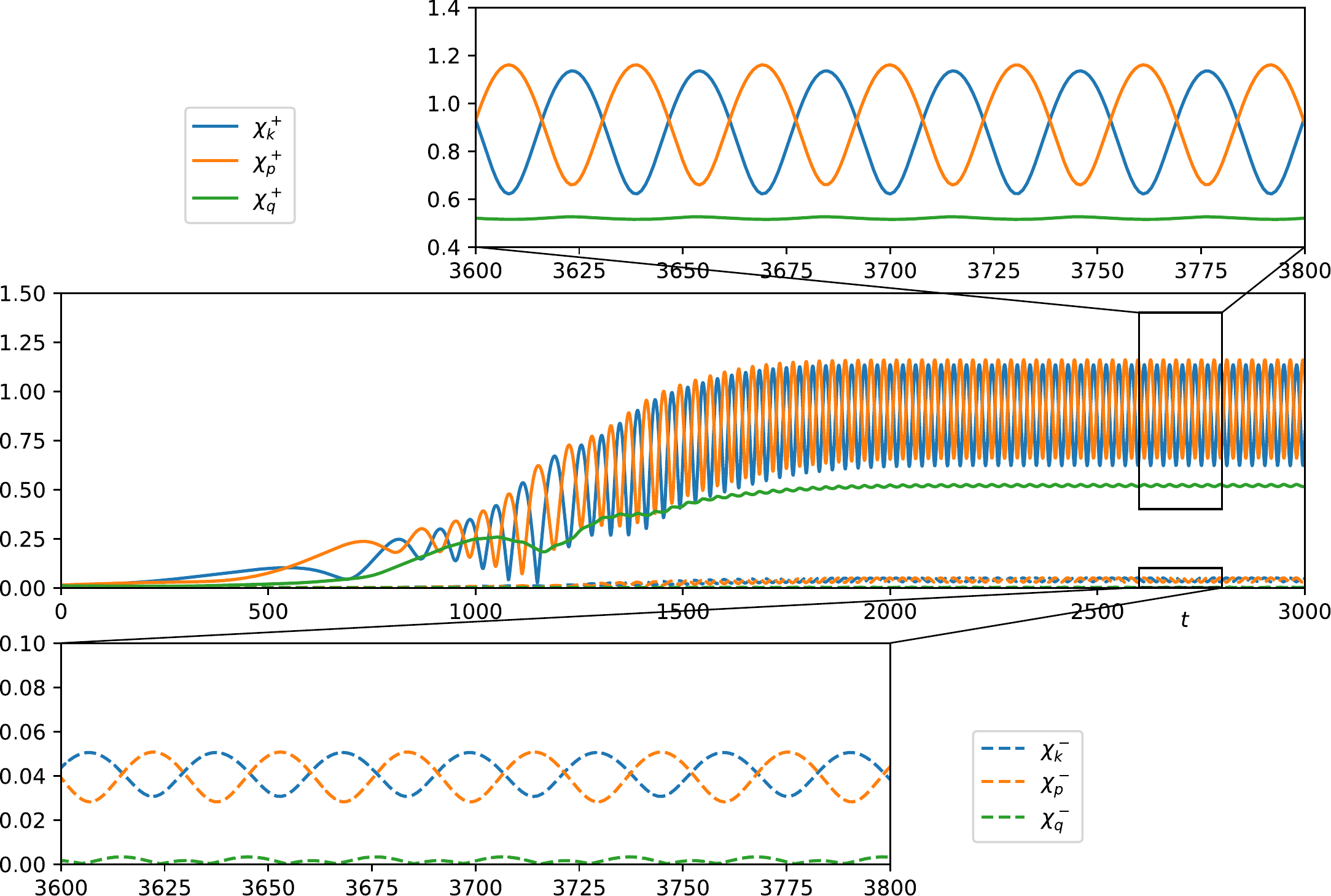}
\par\end{centering}
\caption{\label{fig:amps}Time evolution of the amplitudes of the eigenmodes
for $C=1$ case with $\mathbf{k}=\left(0,1.125\right)$ and $\mathbf{p}=\left(-0.5,-1.0\right)$.
We have a ``saturated'' state with oscillating amplitudes. It seems
that as $k$ and $p$ (the two unstable modes and the two larger legs
of the triads) exchange energy, $q$ plays the role of the mediator.}
\end{figure}

\subsection{Phase Evolution}

Considering the (unwrapped) phase evolution of each of the modes of
the near resonant triad with $\mathbf{k}=\left(0,1.125\right)$ and
$\mathbf{p}=\left(-0.5,-1.0\right)$, we observe that while some amplitude
evolution continues, the phases converge towards straight lines, implying
more or less constant frequencies in the final stage. These nonlinear
frequencies are substantially shifted with respect to the initial
linear frequencies due to the effect of nonlinear terms. However it
appears that the system remains in resonance as the sum of the final
nonlinear frequencies remain very close to zero. In fact, it appears
that the ``near resonant'' system evolves towards resonance as a
result of these nonlinear corrections, since $\Delta\omega$ decreases
from the beginning towards the end.
\begin{figure}[t]
\begin{centering}
\includegraphics[width=1\columnwidth]{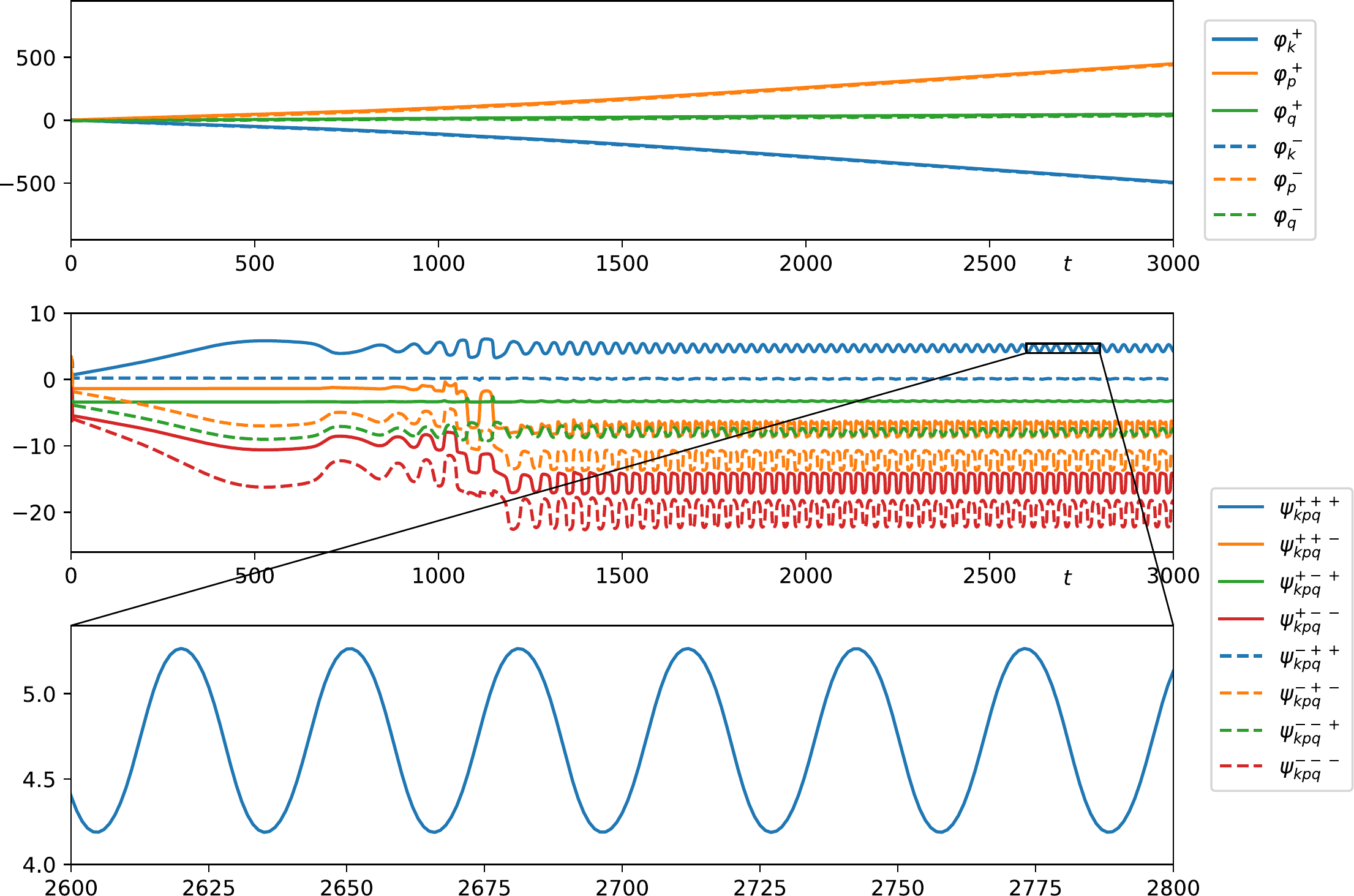}
\par\end{centering}
\caption{\label{fig:phis}Time evolution of the phases $\varphi_{k}^{\pm}$
and their sums $\psi_{kpq}^{s_{p}s_{q}s_{k}}$ for $C=1$ case with
$\mathbf{k}=\left(0,1.125\right)$ and $\mathbf{p}=\left(-0.5,-1.0\right)$.
Saturation of the amplitudes as seen in figure \ref{fig:amps} is
accompanied by a nonlinear frequency shift as shown in the top plot
and the saturation of the $\psi_{kpq}^{s_{p}s_{q}s_{k}}$'s as shown
in the bottom plot. Note that $\psi_{kpq}^{s_{p}s_{q}s_{k}}=\text{const.}$
would correspond to phase locking.}
\end{figure}

Using (\ref{eq:amp}) and (\ref{eq:ph}) with the assumption that
$\partial_{t}\chi_{k}^{\pm}\approx0$ and $\partial_{t}\phi_{k}^{\pm}=-\omega_{k,\text{nl}}^{\pm}$
is a constant, we obtain the nonlinear frequency shift, i.e. $\delta\omega_{kr}^{\pm}=\omega_{k,\text{nl}}^{\pm}-\omega_{kr}^{\pm}$
as: 
\begin{equation}
\delta\omega_{kr}^{\pm}=\text{sign}\left(\omega_{kr}\right)\sqrt{\frac{\left|N_{\xi k}^{\pm}\right|^{2}}{\left|\xi_{k}^{\pm}\right|^{2}}-\gamma_{k}^{\pm2}}\;\text{,}\label{eq:nlfshift}
\end{equation}
which can be computed given the final amplitudes and the nonlinear
interaction coefficients (\ref{NLT}). For example for the case above
the smoothed saturated amplitudes are shown in the table:
\begin{table}[b]
\begin{tabular}{c|cccccc}
\multicolumn{1}{c}{} & $k,+$ & $p,+$ & $q,+$ & $k,-$ & $p,-$ & $q,-$\tabularnewline
\cline{2-7} \cline{3-7} \cline{4-7} \cline{5-7} \cline{6-7} \cline{7-7} 
$\left|\xi\right|$ & $0.89$ & $0.93$ & $0.52$ & $0.041$ & $0.040$ & $0.0017$\tabularnewline
$\omega_{r}$ & $0.099$ & $-0.088$ & $-0.020$ & $-0.099$ & $0.088$ & $0.020$\tabularnewline
$\gamma$ & $4.2\times10^{-3}$ & $3.1\times10^{-3}$ & $-1.8\times10^{-4}$ & $-1.8$ & $-1.8$ & $-4.8$\tabularnewline
$\omega_{\text{nl}}$ & $0.20$ & $-0.19$ & $-0.016$ & $0.20$ & $-0.19$ & $-0.016$\tabularnewline
$\delta\omega$ & $0.11$ & $-0.11$ & \sout{\mbox{$-0.075$}} & \sout{\mbox{$-1.06$}} & \sout{\mbox{$1.12$}} & \sout{\mbox{$23.2$}}\tabularnewline
\end{tabular}

\caption{\label{tab:tablo}Saturated amplitudes, linear frequencies, linear
growth rates, the final nonlinear frequencies and the $\delta\omega$'s
that are computed from (\ref{eq:nlfshift}), rounded to two significant
figures for the $C=1$ case with $\mathbf{k}=\left(0,1.125\right)$
and $\mathbf{p}=\left(-0.5,-1.0\right)$. Note that the basic assumption
of (\ref{eq:nlfshift}) works only for linearly unstable modes, and
for those $\delta\omega$ is not far from $\omega_{\text{nl}}-\omega_{r}$.}

\end{table}

In order to elucidate dynamics of the phases in a triad, we define
the sums of phases as a separate variable following Ref. \citealp{bustamante:09}:
\begin{equation}
\psi_{kpq}^{s_{k}s_{p}s_{q}}\equiv\varphi_{k}^{s_{k}}+\varphi_{p}^{s_{p}}+\varphi_{q}^{s_{q}}\;\text{.}\label{eq:phase_sum}
\end{equation}
We observe that while the phases keep increasing in time, for a steady
state, the phase differences should remain bounded. We can write the
equations for the amplitudes as
\begin{figure}[t]
\begin{centering}
\includegraphics[width=1\columnwidth]{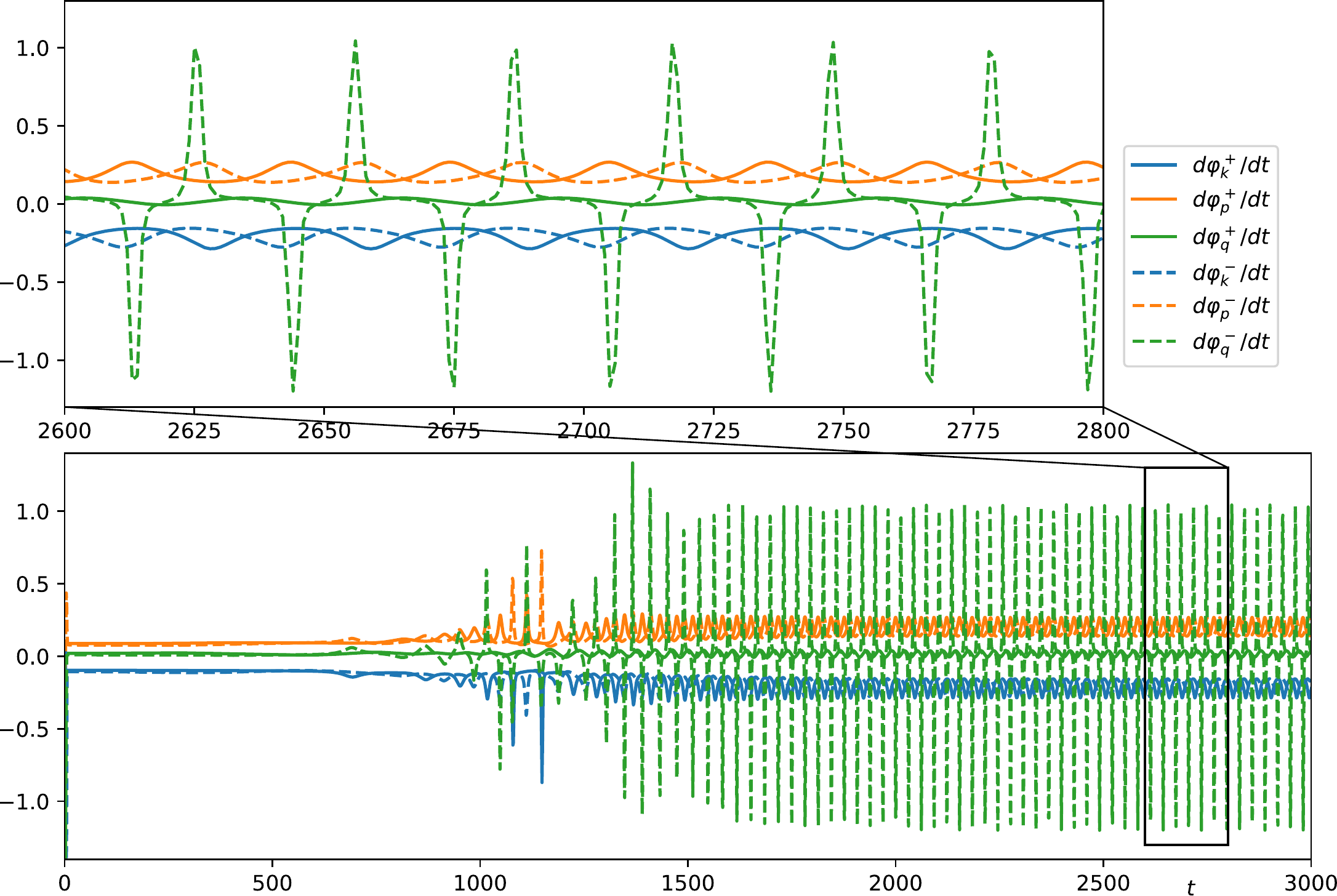}
\par\end{centering}
\caption{\label{fig:phis-1}Time derivatives of the phases $\varphi_{k}^{\pm}$
for $C=1$ case with $\mathbf{k}=\left(0,1.125\right)$ and $\mathbf{p}=\left(-0.5,-1.0\right)$,
corresponding to nonlinear frequencies. Notice that while $d\varphi_{q}^{-}/dt$
appears to oscillate wildly, since its amplitude $\chi_{q}^{-}$ is
vanishingly small, as can be seen in figure \ref{fig:amps}, these
oscillations are not important for the rest of the dynamics.}
\end{figure}
\begin{align}
\partial_{t}\chi_{k}^{s_{k}} & -\gamma_{k}^{s_{k}}\chi_{k}^{s_{k}}\nonumber \\
= & \sum_{\sigma_{p},\sigma_{q}}m_{kpq}^{s_{k}\sigma_{p}\sigma_{q}}\cos\left(\delta_{kpq}^{s_{k}\sigma_{p}\sigma_{q}}-\psi_{kpq}^{s_{k}\sigma_{p}\sigma_{q}}\right)\chi_{p}^{\sigma_{p}}\chi_{q}^{\sigma_{p}}\label{eq:ampsum}
\end{align}
which contain the phases only through their sums (i.e. $\psi$ variables).
We can also write an equation for the $\psi_{kpq}^{s_{k}s_{p}s_{q}}$
explicitly as:
\begin{align}
\partial_{t}\psi_{kpq}^{s_{k}s_{p}s_{q}} & +\text{\ensuremath{\left(\omega_{k}^{s_{k}}+\omega_{p}^{s_{p}}+\omega_{p}^{s_{q}}\right)}}\nonumber \\
 & =\sum_{\sigma_{p},\sigma_{q}}m_{kpq}^{s_{k}\sigma_{p}\sigma_{q}}\sin\left(\delta_{kpq}^{s_{k}\sigma_{p}\sigma_{q}}-\psi_{kpq}^{s_{k}\sigma_{p}\sigma_{q}}\right)\frac{\chi_{p}^{\sigma_{p}}\chi_{q}^{\sigma_{q}}}{\chi_{k}^{s_{k}}}\nonumber \\
 & +\sum_{\sigma_{q},\sigma_{k}}m_{pqk}^{s_{p}\sigma_{q}\sigma_{k}}\sin\left(\delta_{pqk}^{s_{p}\sigma_{q}\sigma_{k}}-\psi_{pqk}^{s_{p}\sigma_{q}\sigma_{k}}\right)\frac{\chi_{q}^{\sigma_{q}}\chi_{k}^{\sigma_{k}}}{\chi_{p}^{s_{p}}}\nonumber \\
 & +\sum_{\sigma_{k},\sigma_{p}}m_{qkp}^{s_{q}\sigma_{k}\sigma_{p}}\sin\left(\delta_{qkp}^{s_{q}\sigma_{k}\sigma_{p}}-\psi_{qkp}^{s_{q}\sigma_{k}\sigma_{p}}\right)\frac{\chi_{k}^{\sigma_{k}}\chi_{p}^{\sigma_{p}}}{\chi_{q}^{s_{q}}}\;\text{.}\label{eq:phsum}
\end{align}
While the form of (\ref{eq:phsum}) looks terribly complicated (e.g.
when we expand the sums we have $8$ equations, each of whom having
$12$ terms on their right hand side) it is useful for insight into
phase locking. For example by setting $\partial_{t}\psi_{kpq}^{s_{k}s_{p}s_{q}}=0$
in (\ref{eq:phsum}), and $\partial_{t}\chi_{k}^{s_{k}}=0$ in (\ref{eq:ampsum}),
we can obtain constant amplitude, phase locked solutions, if such
solutions exist. Unfortunately, even the computation of this ``fixed
point'' requires numerical analysis. We can also integrate (\ref{eq:ampsum}-\ref{eq:phsum})
numerically, which gives exactly the same result as the system in
terms of $\xi_{k}^{\pm}$.

\section{Interactions with Zonal Flows}

When two non-zonal modes interact with a zonal one the evolution equations
and the nonlinear interaction coefficients are different from non-zonal
three wave interactions discussed in the previous section. Using the
original variables $\Phi_{k}$ and $n_{k}$ as in (\ref{eq:hwft1}-\ref{eq:hwft2}),
zonal and non-zonal modes interact with the same nonlinear interaction
coefficients but different linear propagators. However, when we diagonalize
the linear propagator, the nonlinear interaction coefficients for
zonal and non-zonal modes differentiate.

In particular we have 
\begin{equation}
M_{kpq}^{\phi s_{p}s_{q}}=-\frac{\hat{\mathbf{z}}\times\mathbf{p}\cdot\mathbf{q}\left(q^{2}-p^{2}\right)C^{2}}{4\Omega_{p}^{s_{p}*}\Omega_{q}^{s_{q}*}k^{2}p^{2}q^{2}}\label{eq:NLTZk1}
\end{equation}
\begin{align}
M_{kpq}^{ns_{p}s_{q}}=\frac{\hat{\mathbf{z}}\times\mathbf{p}\cdot\mathbf{q}C}{4\Omega_{p}^{s_{p}*}\Omega_{q}^{s_{q}*}p^{2}q^{2}}\bigg[ & \left(B_{q}-i\Omega_{q}^{s_{q}*}\right)q^{2}\nonumber \\
 & -\left(B_{p}-i\Omega_{p}^{s_{p}*}\right)p^{2}\bigg]\label{eq:NLTZk2}
\end{align}
\begin{align}
M_{kpq}^{s_{k}\phi s_{q}}=i\frac{\hat{\mathbf{z}}\times\mathbf{p}\cdot\mathbf{q}}{2\Omega_{q}^{s_{q}*}q^{2}}\bigg[ & \left(B_{q}-i\Omega_{q}^{s_{q}*}\right)q^{2}\nonumber \\
 & -\left(B_{k}-i\Omega_{k}^{s_{k}}\right)\left(q^{2}-p^{2}\right)\bigg]\label{eq:NLTZ1}
\end{align}
\begin{equation}
M_{kpq}^{s_{k}ns_{q}}=i\frac{\hat{\mathbf{z}}\times\mathbf{p}\cdot\mathbf{q}C}{2\Omega_{q}^{s_{q}*}q^{2}}\label{eq:NLTZ2}
\end{equation}
so that for three waves $k$, $p$ and $q$ with $q_{y}=0$, we can
write:
\begin{equation}
\partial_{t}\overline{\Phi}_{q}+\nu_{ZF}\overline{\Phi}_{q}=\sum_{s_{k},s_{p}}M_{qkp}^{\phi s_{k}s_{p}}\xi_{k}^{s_{k}*}\xi_{p}^{s_{p}*}\label{eq:zf1}
\end{equation}
\begin{equation}
\partial_{t}\overline{n}_{q}+D_{ZF}\overline{n}_{q}=\sum_{s_{k},s_{p}}M_{qkp}^{ns_{k}s_{p}}\xi_{k}^{s_{k}*}\xi_{p}^{s_{p}*}\label{eq:zf2}
\end{equation}
\begin{equation}
\partial_{t}\xi_{k}^{s_{k}}+i\omega_{k}^{s_{k}}\xi_{k}^{s_{k}}=\sum_{s_{p}}M_{kpq}^{s_{k}s_{p}\phi}\xi_{p}^{s_{p}*}\Phi_{q}^{*}+M_{kpq}^{s_{k}s_{p}n}\xi_{p}^{s_{p}*}n_{q}^{*}\label{eq:xizf1}
\end{equation}
\begin{equation}
\partial_{t}\xi_{p}^{s_{p}}+i\omega_{p}^{s_{p}}\xi_{p}^{s_{p}}=\sum_{s_{p}}M_{pqk}^{s_{p}\phi s_{k}}\Phi_{q}^{*}\xi_{k}^{s_{k}*}+M_{pqk}^{s_{p}ns_{k}}n_{q}^{*}\xi_{k}^{s_{k}*}\;\text{.}\label{eq:xizf2}
\end{equation}
We can write these in the form (\ref{eq:xik}-\ref{eq:xiq}) by letting
$\xi_{q}^{+}=\Phi_{q}$ and $\xi_{q}^{-}=n_{q}$ and paying attention
to the form of the interaction coefficient $M_{\xi kpq}^{s_{k}s_{p}s_{q}}$
when one of the legs is zonal.

\begin{figure}
\begin{centering}
\includegraphics[width=1\columnwidth]{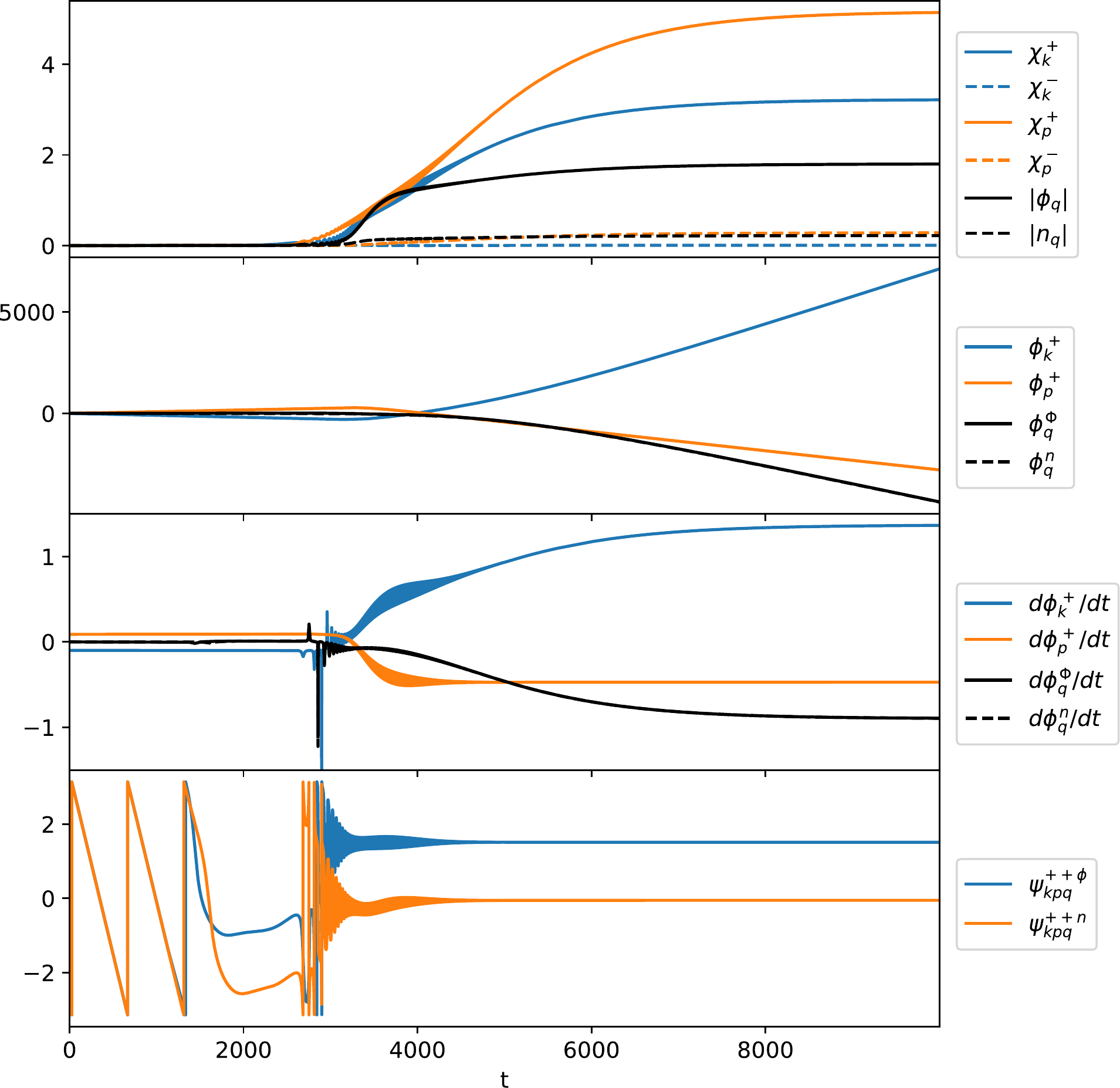}
\par\end{centering}
\caption{\label{fig:zfsup}Time evolution of the three wave equations involving
a zonal mode $q$, for the case $C=1$, $\nu_{Z}=D_{Z}=0$ and $\gamma_{k}\gtrsim\gamma_{p}>0$
with $k_{y}=1.125$ and $q=0.5$ {[}i.e. $\mathbf{k}=\left(0,k_{y}\right)$,
$\mathbf{p}=\left(-q,-k_{y}\right)$ and $\mathbf{q}=\left(q,0\right)${]}.
The system reaches a steady state by introducing nonlinear frequencies
in order to arrive at a state where the sums of phases $\psi_{kpq}$'s
are constant. Note that it is $p$, which becomes the dominant mode
in the final state and the existence of zonal flows does not lead
to a complete suppression of turbulence. Instead the zonal flow acquires
a constant nonlinear frequency.}
\end{figure}

In order to study the interactions between two modes with a zonal
flow in the Hasegawa-Wakatani system numerically, we pick a primary
wave-number $\mathbf{k}=\left(0,1.125\right)$ which is the linearly
most unstable mode on a grid with $dk_{x}=dk_{y}=0.125$ for the $C=1.0$
case. We choose $\mathbf{p}=\left(-0.5,-1.125\right)$ so that $\mathbf{q}=\left(0.5,0\right)$
is a zonal wave number. The $6$ field variables are now $\xi_{k}^{\pm}$,
$\xi_{p}^{\pm}$, $\Phi_{q}$ and $n_{q}$ whose evolutions are shown
in figure \ref{fig:zfsup} for the case $C=1$, $\nu_{Z}=D_{Z}=0$
and $\gamma_{k}\gtrsim\gamma_{p}>0$. In the final state, the system
finds a fixed point characterized by constant nonlinear frequency
shifts, constant amplitudes and constant $\psi_{kpq}$'s. However
this kind of steady state solution seems to be exclusive to the single
triad case.

\subsection{Triad pairs\label{subsec:Triad-pairs}}

Because of the symmetry of the system, if we consider two wave-numbers
$\mathbf{p}_{1}=-\mathbf{k}-\mathbf{q}$ and $\mathbf{p}_{2}=-\mathbf{k}+\mathbf{q}$
with $\mathbf{k}$ in $\hat{\mathbf{y}}$ and $\mathbf{q}$ in $\hat{\mathbf{x}}$
directions, we get two triads that are reflections of one another
with respect to the axis defined by $\mathbf{k}$. Such a system involves
four different wave-numbers connected with two different triads. Including
the $p\rightleftarrows q$ transformation we have four triads as shown
in figure \ref{fig:triad4}. However as long as we use symmetric forms
for the interaction coefficients, we can drop the two triads we obtain
from the $p\rightleftarrows q$ transformation and count only two
triads. Since the two triads of such a pair are reflections of one
another, the nonlinear interaction coefficients differ only in sign
while the complex frequencies are the same, and as there are two eigenmodes
for each wave-number, we have $8$ equations. The equations for zonal
modes can be written from (\ref{eq:zf1}-\ref{eq:zf2}) as:
\begin{figure}
\begin{centering}
\includegraphics[width=1\columnwidth]{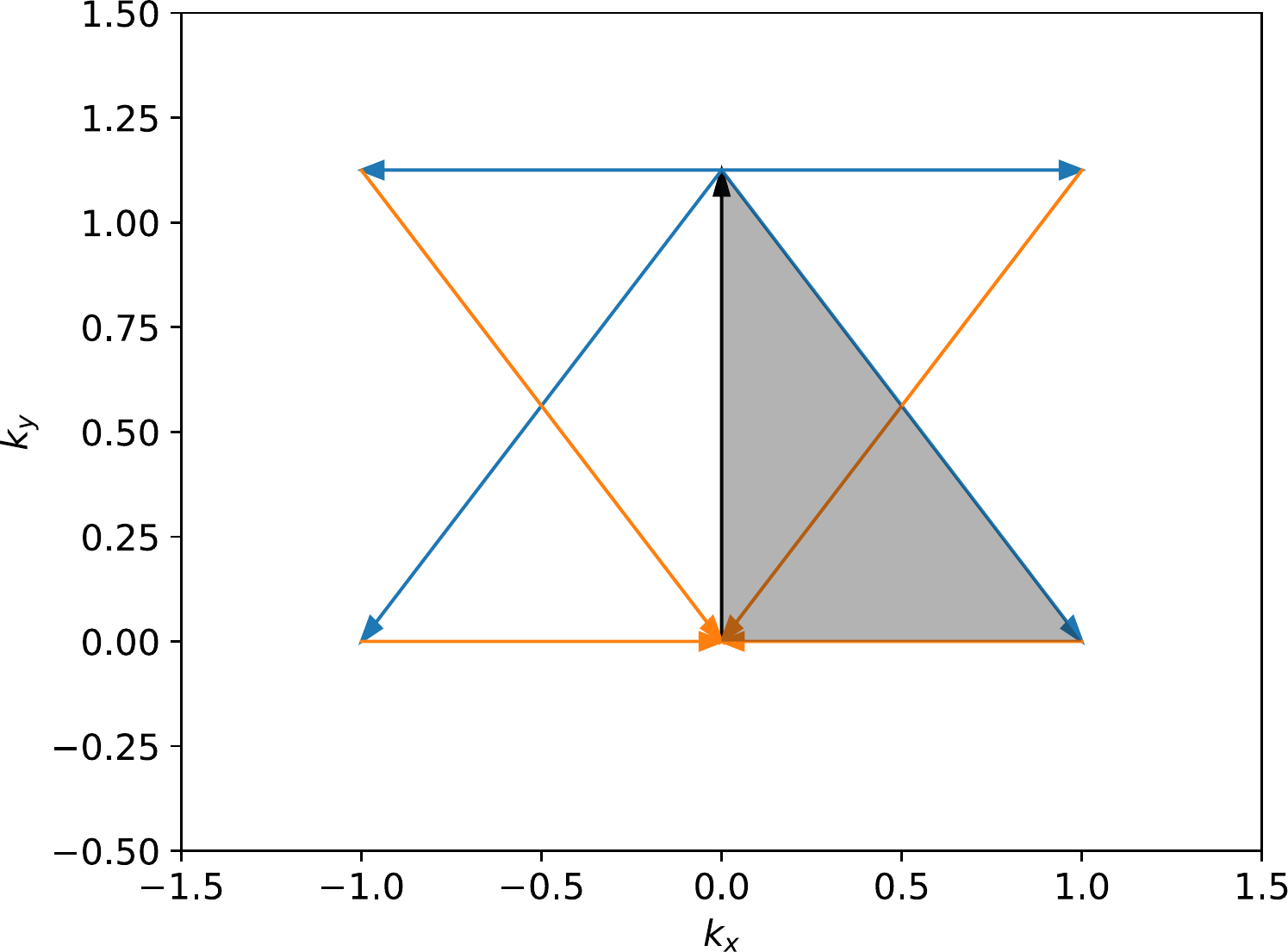}
\par\end{centering}
\caption{\label{fig:triad4} All the four triads involved in the interaction
between the most unstable mode with $\mathbf{k}=k_{y}\hat{\mathbf{y}}$
with $k_{y}=1.125$ and a given zonal mode with $q=1.0$, obtained
by reflection with respect to $\mathbf{k}$ and the exchange of $p$
and $q$ of the primary triad, which is shaded. The existence of the
reflected triad is indeed important as it changes the qualitative
behavior with respect to the single triad case.}
\end{figure}
\begin{equation}
\partial_{t}\Phi_{q}+\nu_{Z}\Phi_{q}=\sum_{s_{k},s_{p}}M_{qkp_{1}}^{\phi s_{k}s_{p}}\left(\xi_{k}^{s_{k}*}\xi_{p_{1}}^{s_{p_{1}}*}-\xi_{k}^{s_{k}}\xi_{p_{2}}^{s_{p_{2}}}\right)\label{eq:4wzf1}
\end{equation}
\begin{equation}
\partial_{t}n_{q}+D_{Z}n_{q}=\sum_{s_{k},s_{p}}M_{qkp}^{ns_{k}s_{p}}\left(\xi_{k}^{s_{k}*}\xi_{p_{1}}^{s_{p}*}-\xi_{k}^{s_{k}}\xi_{p_{2}}^{s_{p}}\right)\;\text{,}\label{eq:4wzf2}
\end{equation}
which is possible since $M_{\xi kp_{2}q}^{s_{k}s_{p}\left\{ n,\phi\right\} }=-M_{\xi kp_{1}q}^{s_{k}s_{p}\left\{ n,\phi\right\} }$
because $p_{2}^{2}=p_{1}^{2}$ and $p_{2y}=p_{1y}$ while $p_{2x}=-p_{2x}$.
The equation for the primary mode, can be written as:
\begin{align}
\partial_{t}\xi_{k}^{s_{k}}+i\omega_{k}^{s_{k}}\xi_{k}^{s_{k}} & =\sum_{s_{p}}\bigg[M_{\xi kpq}^{s_{k}s_{p}\phi}\left(\Phi_{q}^{*}\xi_{p_{1}}^{s_{p}*}+\Phi_{q}\xi_{p_{2}}^{s_{p}*}\right)\nonumber \\
 & +M_{\xi kpq}^{s_{k}s_{p}n}\left(n_{q}^{*}\xi_{p_{1}}^{s_{p}*}+n_{q}\xi_{p_{2}}^{s_{p}*}\right)\bigg]\;\text{,}\label{eq:4wzf3}
\end{align}
and the remaining two equations are the same as (\ref{eq:xizf2})
but with different signs and conjugations: 
\begin{equation}
\partial_{t}\xi_{p_{1}}^{s_{p}}+i\omega_{p_{1}}^{s_{p}}\xi_{p_{1}}^{s_{p}}=\sum_{s_{k}}\left(M_{\xi p_{1}kq}^{s_{p}s_{k}\phi}\Phi_{q}^{*}+M_{\xi p_{1}kq}^{s_{p}s_{k}n}n_{q}^{*}\right)\xi_{k}^{s_{k}*}\label{eq:4wzf4}
\end{equation}
\begin{equation}
\partial_{t}\xi_{p_{2}}^{s_{p}}+i\omega_{p_{2}}^{s_{p}}\xi_{p_{2}}^{s_{p}}=-\sum_{s_{k}}\left(M_{\xi p_{1}kq}^{s_{p}s_{k}\phi}\Phi_{q}+M_{\xi p_{1}kq}^{s_{p}s_{k}n}n_{q}\right)\xi_{k}^{s_{k}*}\label{eq:4wzf5}
\end{equation}
where $\omega_{p_{2}}^{s_{p}}=\omega_{p_{1}}^{s_{p}}$. Notice that
this is also equivalent to one of the radial Fourier modes of a quasi-linear
(e.g. zonostrophic) interaction, where for each field one would consider
a single $p_{y}$ but the full spatial dependence in $x$.
\begin{figure}
\begin{centering}
\includegraphics[width=1\columnwidth]{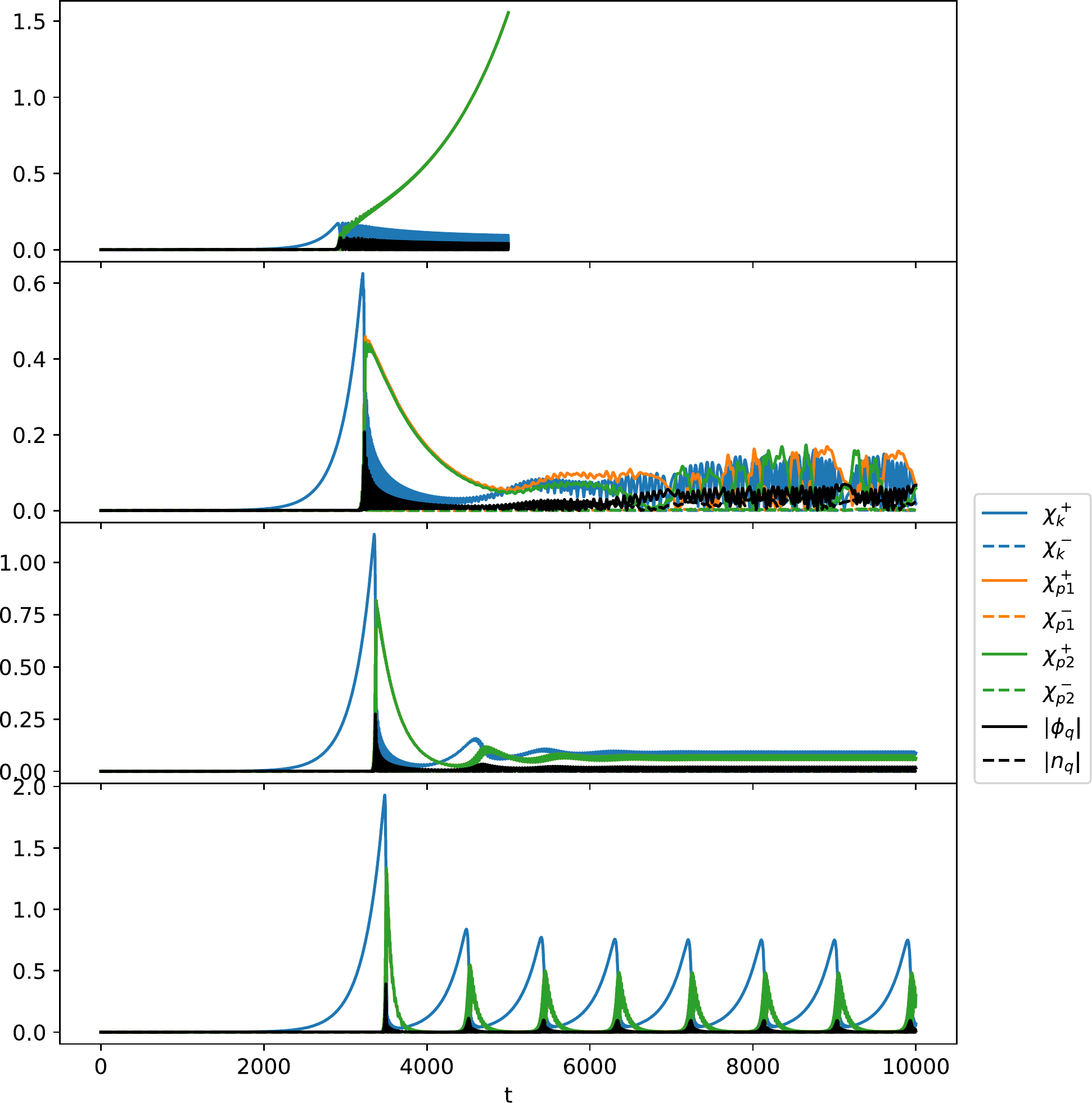}
\par\end{centering}
\caption{\label{fig:zfamps3}Evolution of a triad pair with the same parameters
as figure \ref{fig:zfsup}, no zonal flow damping $\nu_{Z}=D_{Z}=0$
and $k_{y}=1.125$ {[}i.e. $\mathbf{k}=\left(0,k_{y}\right)$, $\mathbf{p}_{1}=\left(-q,-k_{y}\right)$,
$\mathbf{p}_{2}=\left(q,-k_{y}\right)$ and $\mathbf{q}=\left(q,0\right)${]}
for four different values of $q=\left(1.0,1.5,2.0,4.0\right)$ from
top to bottom for which the growth rates of the subdominant modes
are $\gamma_{p}=\left(0.00099,-0.0016,-0.0042,-0.017\right)$ respectively.
Note that apart from the second plot, which displays some chaotic
behavior, the curves for $\xi_{p_{1}}^{+}$ and $\xi_{p_{2}}^{+}$
overlap almost exactly.}
\end{figure}

The results of the system (\ref{eq:4wzf1}-\ref{eq:4wzf5}) are shown
in figure \ref{fig:zfamps3} for the $C=1$ case with $k_{y}=1.125$
{[}i.e. $\mathbf{k}=\left(0,k_{y}\right)$, $\mathbf{p}_{1}=\left(-q,-k_{y}\right)$,
$\mathbf{p}_{2}=\left(q,-k_{y}\right)$ and $\mathbf{q}=\left(q,0\right)${]}
for $q=\left(1.0,1.5,2.0,4.0\right)$ from top to bottom respectively.
For $q\leq k_{y}$ we have instability and $p$ keeps growing exponentially
whereas for $q>k_{y}$ we get some sort of steady or limit cycle state.
Performing a scan of $k_{y}$ and $q$ for this two triad system (keeping
in mind that for $k_{y}>2$ we have no instability and therefore the
pump mode decays) we observe that we can define a four wave interaction
condition of the form $\omega_{kr}^{s_{k}}+\omega_{p1r}^{s_{p1}}+\omega_{p2r}^{s_{p2}}+\omega_{qr}^{s_{q}}=0$,
which turns into $\Omega_{k}^{s_{k}}+2\Omega_{p}^{s_{p}}=0$ since
$\omega_{qr}=0$ and $\omega_{p1r}=\omega_{p2r}=\omega_{pr}$. There
seems to be $3$ distinct regions in figure X: for $q<1$, the $\xi_{p}^{+}$modes
grow exponentially as in the top plot of figure \ref{fig:zfamps3},
for the central region where $q\approx1$, we have saturation and
then somewhat chaotic evolution, and finally for $q\gg1$, we observe
limit cycle oscillations between $\xi_{k}^{+}$ and $\xi_{p}^{+}$
modes, mediated by zonal flows.

One is tempted to argue that since the $p$ with $p_{x}<p_{y}$ wins
the competition to attract more energy, the cascade will proceed in
this direction, and in the next step we can consider the interaction
of this $\xi_{p}^{+}$ as the pump mode for the next triad etc. However,
since each mode interacts with many triads simultaneously, the fact
that $\xi_{p}^{+}$ wins the competition in the single triad (or one
triad and its reflection) configuration does not really mean the energy
will indeed go this way. 

\subsection{Triad Networks}

In order to study the fate of the cascade, we need to consider multiple
triads that are connected to one another. However as we add more zonal
and non-zonal modes, it becomes quite complicated to keep track of
all the interaction coefficients, conjugations etc. In order to simplify
this task, we can divide the problem into two steps i) construction
of a network of three body interactions and ii) computation of the
evolution of the field variables on this network. For example for
the above problem we need to consider a network of $N_{k}=4$ wave
number nodes, coupled to $N_{t}=2$ triads, with $N_{f}=2$ fields
in each node, with an interaction coefficient of the size $N_{f}\times N_{f}\times N_{f}$
for each connection. Since a network in Fourier space is made up of
three body interactions, for each node, we can compute a list of interacting
pairs and the interaction coefficients, so that we can write
\begin{equation}
\partial_{t}\xi_{\ell}^{i}+L_{\ell}^{ij}\xi_{\ell}^{j}=\frac{1}{2N}\sum_{\ell',\ell''=\mathbf{i}_{\ell}}M_{\ell\ell'\ell''}^{ijk}\left(\xi_{\ell'}^{j}\right)^{c_{\ell'}}\left(\xi_{\ell''}^{k}\right)^{c_{\ell''}}\label{eq:nw1}
\end{equation}
where $\mathbf{i}_{\ell}$ is the list of precomputed interaction
pairs for the node $\ell$. The indices $i$, $j$ and $k$ correspond
to different fields (eigenmodes or $\Phi_{k}$ and $n_{k}$), the
matrix $L_{\ell}^{ij}$ is the linear matrix in $k$ space (i.e. diagonal
with the elements $i\omega_{\ell}^{\pm}$ for the eigenmodes), the
$M_{\ell\ell'\ell''}^{ijk}$ is the interaction coefficient for each
interaction and $N$ is the number of independent wave number nodes
so that when we reach the full grid, we have exactly the same interaction
coefficients as the system formulated using discrete fast Fourier
transforms (i.e. divided by $N_{x}\times N_{y}$). Finally if we write
the triad interaction condition in the form $\mathbf{k}_{\ell}+\sigma_{\ell'}\mathbf{k}_{\ell'}+\sigma_{\ell''}\mathbf{k}_{\ell''}=0$
, where $\sigma$ are $\pm1$, the $\left(\xi_{\ell'}^{j}\right)^{c_{\ell'}}$
are defined as:
\begin{figure}
\begin{centering}
\includegraphics[width=1\columnwidth]{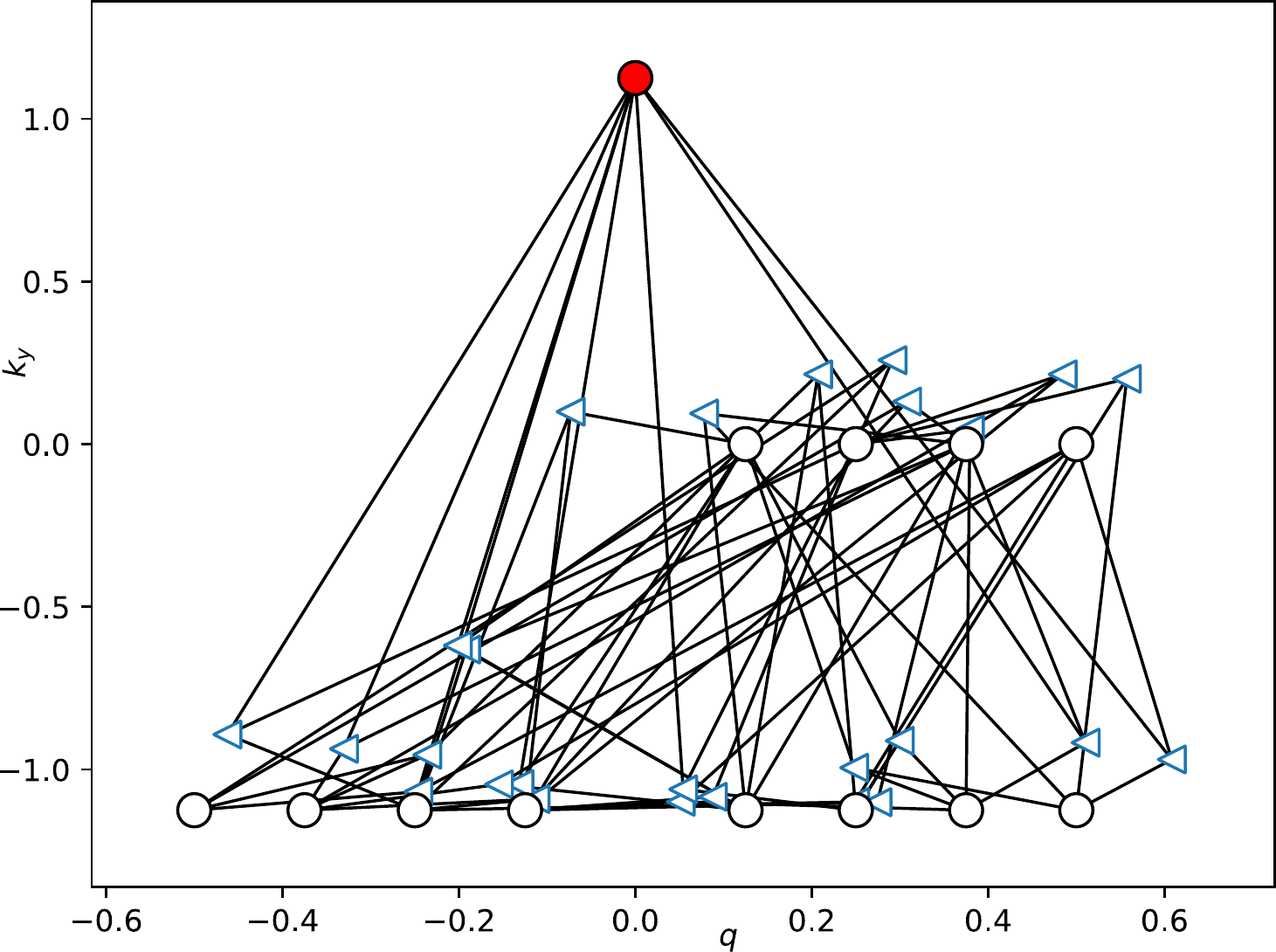}
\par\end{centering}
\caption{\label{fig:nw1}The structure of the network with a single $k_{y}$
with $k_{y}=1.125$ shown as a filled (red if in color) node. A reduced
version with $q$ values that only go up to $0.5$ is shown for clarity.
Notice that in this network while all of the $26$ triads involve
one of the zonal modes, only $8$ of them involve the $q=0$ mode.}
\end{figure}
\[
\left(\xi_{\ell'}^{j}\right)^{c_{\ell'}}=\begin{cases}
\xi_{\ell'}^{j} & \sigma_{\ell'}=-1\\
\xi_{\ell'}^{j*} & \sigma_{\ell'}=+1
\end{cases}
\]
This is necessary unless we have the negative of each wave number
vector as a separate node in the network.

Notice that when computing the nonlinear interaction coefficients
for the eigenmodes, we would use (\ref{NLT}) if all the nodes have
nonzero $k_{y}$. In contrast we would use (\ref{eq:NLTZk1}) and
(\ref{eq:NLTZk2}) if the receiving node (i.e. node $\ell$ ) is zonal
or (\ref{eq:NLTZ1}) and (\ref{eq:NLTZ2}) if one of the interacting
pairs (i.e. $\ell'$ or $\ell''$) are zonal. Two or more zonal mode
do not interact because of the geometric factor $\hat{\mathbf{z}}\times\mathbf{p}\cdot\mathbf{q}$,
which appear in front of all the interaction coefficients.

Finally, if it makes sense to zero out some of the fields at a given
wave-number (e.g. in eigenmode formulation, we may decide to throw
away some damped modes), one may switch to a formulation where each
node corresponds to a wave-number/field variable combination via $\left\{ k_{x},k_{y},s_{k}\right\} \rightarrow\ell$.
In this case, assuming that the linear matrix $L_{\ell}^{ij}$ in
(\ref{eq:nw1}) diagonal takes the form:
\begin{equation}
\partial_{t}\xi_{\ell}+i\omega_{\ell}\xi_{\ell}=\frac{1}{N}\sum_{\ell',\ell''=\mathbf{i}_{\ell}}M_{\ell\ell'\ell''}\xi_{\ell'}^{c_{\ell'}}\xi_{\ell''}^{c_{\ell''}}\label{eq:nwsimp}
\end{equation}

\subsection{Order Parameters}

The phases of wave-number nodes in Hasegawa-Wakatani turbulence evolve
according to (\ref{eq:ph}) or written explicitly as (\ref{eq:phase}).
This suggests that one can possibly define some kind of order parameter
for this system. The usual definition of the Kuramoto order parameter
can be written for the network formulation of (\ref{eq:nwsimp}) as:
\begin{equation}
z=re^{i\psi}=\frac{1}{N}\sum_{\ell}e^{i\varphi_{\ell}}\label{eq:z_kuro}
\end{equation}
without explicitly distinguishing $+$ or $-$ modes. However this
order parameter based on an unweighted sum is probably relevant only
if all the oscillators were identical with all-to-all, unweighted
couplings of the Kuramoto type. Instead we can use an amplitude filtered
Kuramoto order parameter (i.e. the sum is computed only over the oscillators
with an amplitude larger than a threshold), or define a weighted version
of (\ref{eq:z_kuro}) as:
\begin{equation}
z=re^{i\psi}=\frac{\sum_{\ell}\chi_{\ell}e^{i\varphi_{\ell}}}{\sum_{\ell}\chi_{\ell}}\label{eq:z_weg}
\end{equation}
whose absolute value would tends towards $1$ if the relevant phases
(i.e. those that have large amplitude) are the same. However note
that the weighted order parameter tends towards $1$ also when one
of the modes dominate over the others, while $\psi$ as defined in
(\ref{eq:z_weg}), can still be used as a mean phase.

\begin{figure}
\begin{centering}
\includegraphics[width=1\columnwidth]{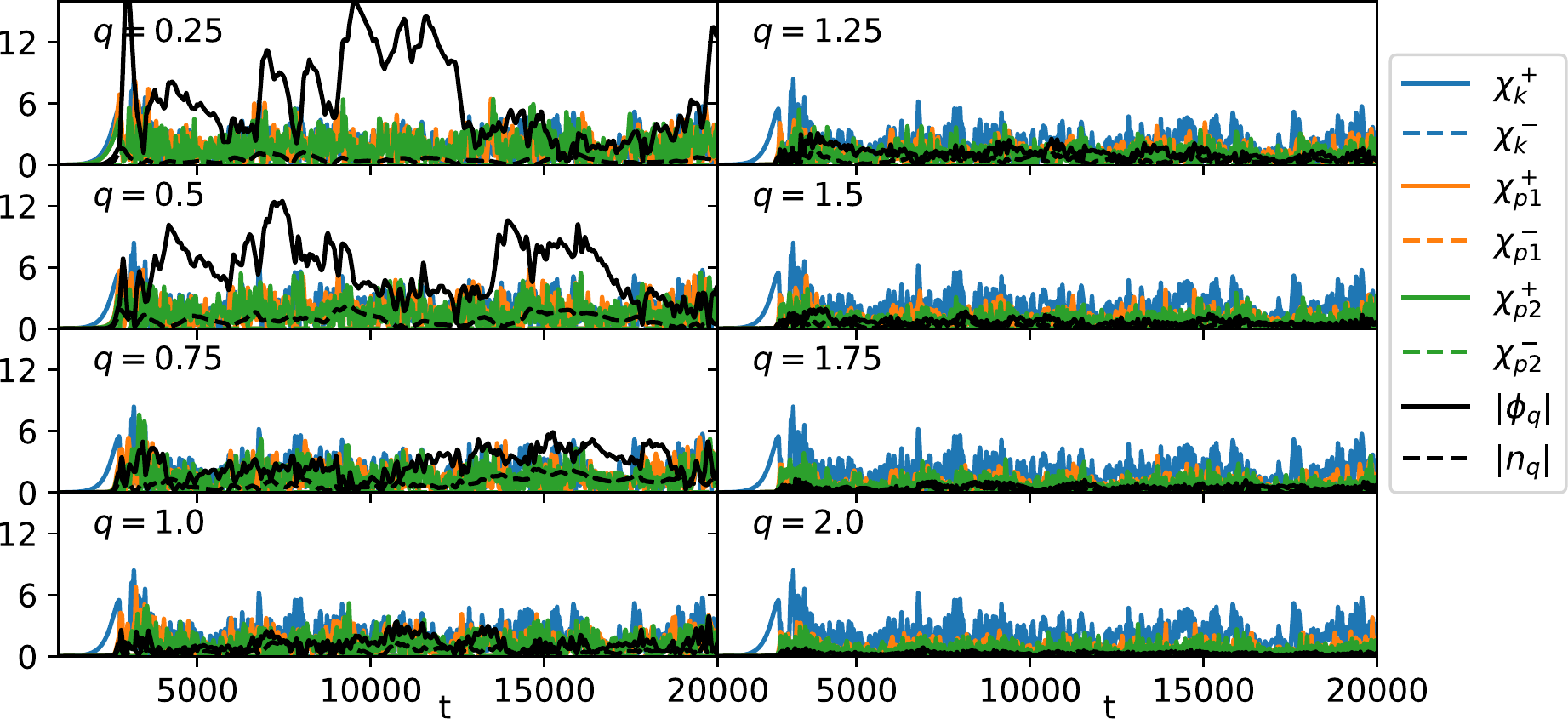}
\par\end{centering}
\caption{\label{fig:nwts2} Time evolution for a number of triad pairs (as
defined in section \ref{subsec:Triad-pairs}) with different values
of $q$ in the network of interacting triads for $C=1$ case with
$\nu_{ZF}=D_{ZF}=10^{-3}$. A steady state turbulence level is observed,
with elevated levels of zonal flows at large scales.}
\end{figure}

It would also make sense to look at the net effect on the nonlinear
term on the phases instead. As discussed in Section \ref{subsec:Amplitude-and-Phase},
since we can write:
\begin{equation}
\partial_{t}\varphi_{\ell}=-\omega_{\ell}+\frac{1}{N\chi_{\ell}}\text{Im}\left[\sum_{\ell',\ell''=\mathbf{i}_{\ell}}M_{\ell\ell'\ell''}\xi_{\ell'}^{c_{\ell'}}\xi_{\ell''}^{c_{\ell''}}e^{-i\varphi_{\ell}}\right]\label{eq:ph2}
\end{equation}
for the evolution of the phase, we can define:
\begin{equation}
Z_{\ell}=R_{\ell}e^{i\psi_{\ell}}=\frac{1}{N\chi_{\ell}}\left(\sum_{\ell',\ell''=\mathbf{i}_{\ell}}M_{\ell\ell'\ell''}\xi_{\ell'}^{c_{\ell'}}\xi_{\ell''}^{c_{\ell''}}\right)\label{eq:zl}
\end{equation}
with $d_{\ell}$ being the number of interactions for the node $\ell$
(i.e. length of $\mathbf{i}_{\ell}$), as some kind of local order
parameter for the node $\ell$, allowing us to write the phase equation
as:
\begin{equation}
\partial_{t}\varphi_{\ell}=-\omega_{\ell}+R_{\ell}\sin\left(\psi_{\ell}-\varphi_{\ell}\right)\;\text{,}\label{eq:ph_simp}
\end{equation}
which attracts the system towards $\varphi_{\ell}=\psi_{\ell}+2n\pi$.
\begin{figure}
\begin{centering}
\includegraphics[width=1\columnwidth]{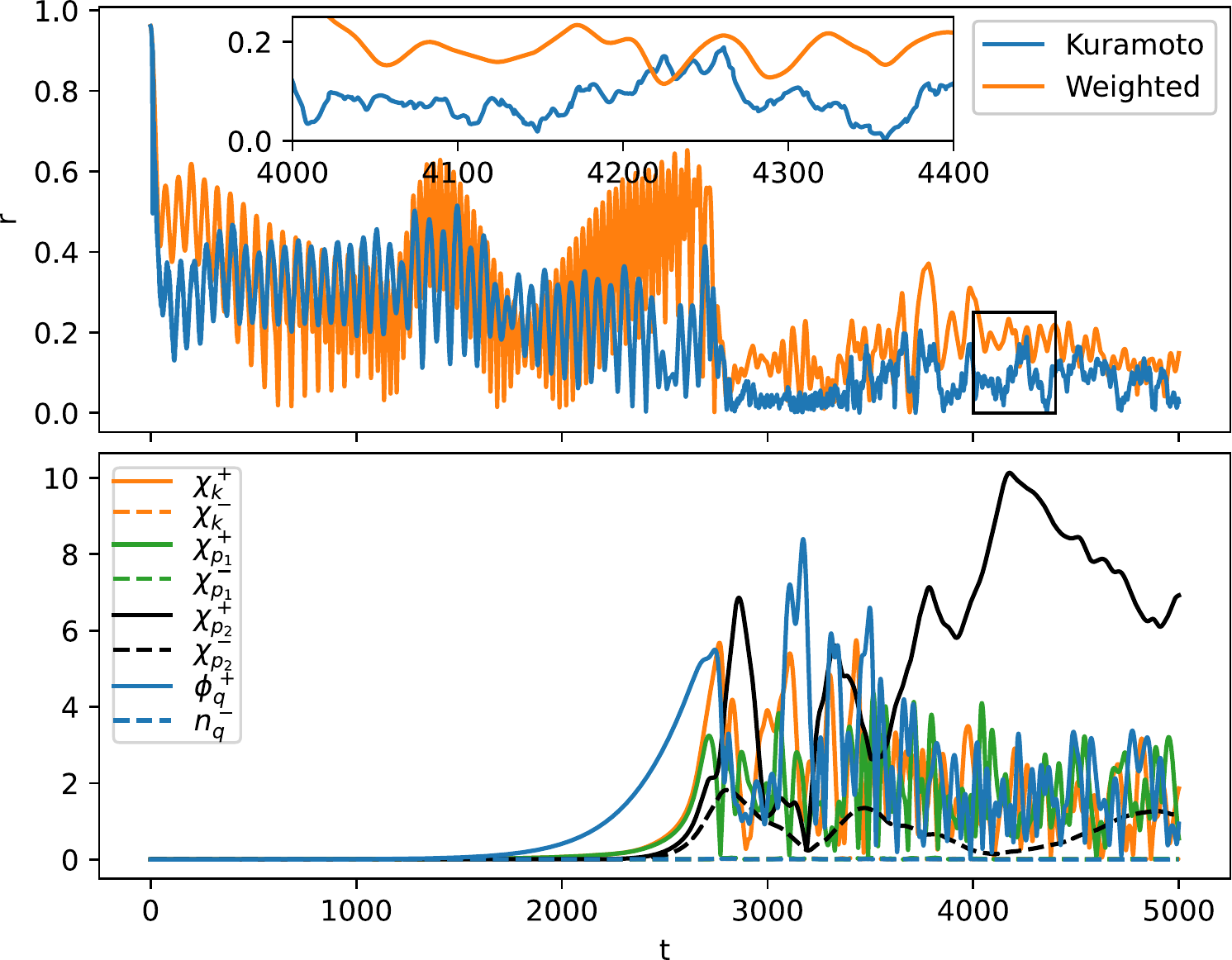}
\par\end{centering}
\caption{\label{fig:order2-1}The top plot shows the order parameter $r$ defined
in (\ref{eq:z_kuro}) or (\ref{eq:z_weg}) as a function of time for
a network with single $q$ and multiple $k_{y}$. The two definitions
are in reasonable agreement apart from the peak around $t=2500$ for
the weighted order parameter, which corresponds to the linear growth
phase, where only a few modes around the most unstable mode dominate.
This can be seen at the bottom plot where the amplitudes of a triad
pair with $q=0.5$ and $k_{y}=1.125$ are shown. Around $t=2500$
the blue curve clearly dominates.}
\end{figure}

\subsection{Specific network configurations}

\subsubsection{Network with a single $k_{y}$:}

We consider a network of triad pairs as discussed in section \ref{subsec:Triad-pairs},
with a single value of $k_{y}$ and $q$ values that go from $0.125$
to $4.0$ in steps of $0.125$. Notice that such a network has many
different types of interactions as shown in figure \ref{fig:nw1},
but all of those involve one of the zonal modes, which means that
if we compute the inverse Fourier transform in the $x$ direction,
the network can be seen to be equivalent to the single $k_{y}$, full-$x$,
quasi-linear model \citep{bian:03,sarazin:21}, since in both cases
we have full spatial evolution but only nonlinear coupling is with
the zonal flow.
\begin{figure}
\begin{centering}
\includegraphics[width=1\columnwidth]{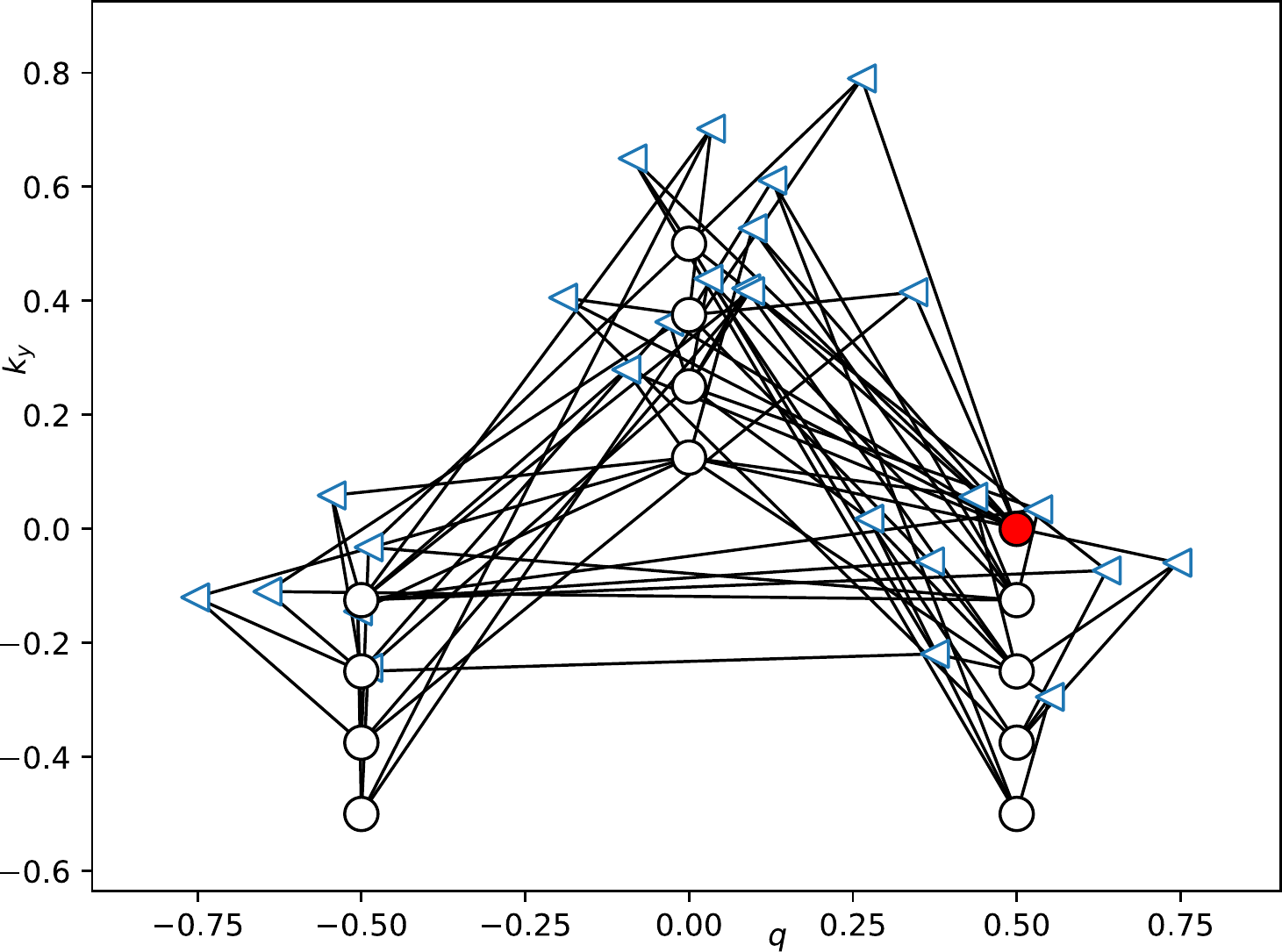}
\par\end{centering}
\caption{\label{fig:nw2}The structure of the network with a single $q=0.5$
zonal mode, shown as a filled (red if in color) node. A reduced version
with $k_{y}$ values that only go up to $0.5$ is shown for clarity.
Only $8$ of the full $26$ triads involve the zonal flow.}
\end{figure}

For the case $C=1$, without zonal flow damping (not shown)  we observe
that the zonal flows dominate and all the other modes decay to zero.
This may well be what happens also in direct numerical simulations
(DNS) eventually: what we observe in numerical simulations without
zonal flow damping is a continual increase of zonal flows even for
very long simulations. 

In contrast, when we introduce zonal flow damping by letting $\nu_{ZF}=D_{ZF}=10^{-3}$,
we get dynamics and $k$-spectra which look more like fully developed
Hasegawa-Wakatani turbulence, as shown in figure \ref{fig:nwts2},
with high levels of zonal flows at large scales.
\begin{figure}
\begin{centering}
\includegraphics[width=1\columnwidth]{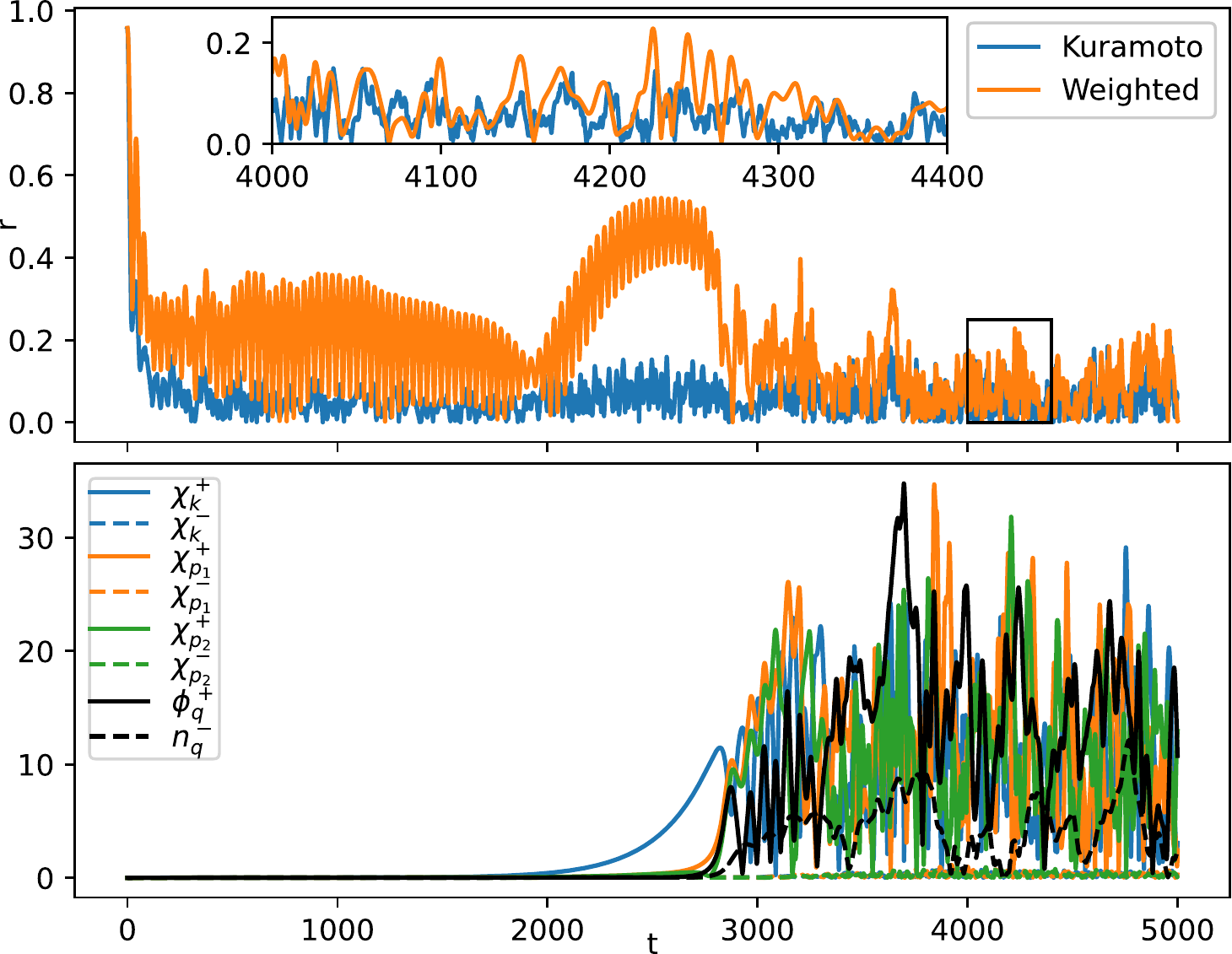}
\par\end{centering}
\caption{\label{fig:order2}The top plot shows the order parameter $r$ defined
in (\ref{eq:z_kuro}) or (\ref{eq:z_weg}) as a function of time for
a network with single $q$ and multiple $k_{y}$. The two definitions
are in reasonable agreement apart from the peak around $t=2500$ for
the weighted order parameter, which corresponds to the linear growth
phase, where only a few modes around the most unstable mode dominate.
This can be seen at the bottom plot where the amplitudes of a triad
pair with $q=0.5$ and $k_{y}=1.125$ are shown. Around $t=2500$
the blue curve clearly dominates.}
\end{figure}

\subsubsection{Network with a single $q$:}

Here, we consider a network of triad pairs with a single $q$, and
a grid of values of $k_{y}$ going from $0.125$ to $4.0$ in steps
of $0.125$. A reduced version of such a network is shown in figure
\ref{fig:nw2}. Physically this network corresponds to the opposite
case where we consider a single $q$ with the whole $y$ dynamics
if we compute the inverse Fourier transform in $y$. Since it involves
bunch of oscillators with different frequencies (as $\omega$ is mostly
a function of $k_{y}$) that are coupled to each other and to a zonal
mode that may play the role of a dominant mean field, it has the basic
ingredients that may lead to synchronization.

Nonetheless numerical observations suggest that there is no obvious
route to global synchronization in the three body network of interacting
triads consisting of a zonal mode and drift waves of different $k_{y}$
either. The weighted order parameter shows a brief increase during
the nonlinear saturation phase as the energy is transferred to the
zonal flow, but otherwise remain close to zero, while the Kuramoto
order parameter simply remains close to zero the whole time as can
be seen in figure \ref{fig:order2}. Since we observed no qualitative
difference between the runs with or without zonal flow damping for
this case, we only show those with $\nu_{ZF}=D_{ZF}=10^{-3}$.

\subsection{Direct numerical simulations}

One can think of direct numerical simulation (DNS) on a regular rectangular
grid as a ``network'' in Fourier space, in the sense that it consists
of a collection of wave number nodes connected to each other through
triadic interactions. In contrast to the networks that we considered
that contain a single zonal mode, or a single $q=0$ mode, a regular
rectangular grid has all the possible wave-numbers in a particular
range, and it allows using more efficient methods for computing the
convolution sums. In practice, the high resolution direct numerical
simulations that we discuss here were performed with a standard pseudo-spectral
solver (i.e. with periodic boundary conditions in both directions)
using 2/3 rule for dealiasing and adaptive time stepping.
\begin{figure}
\begin{centering}
\includegraphics[width=1\columnwidth]{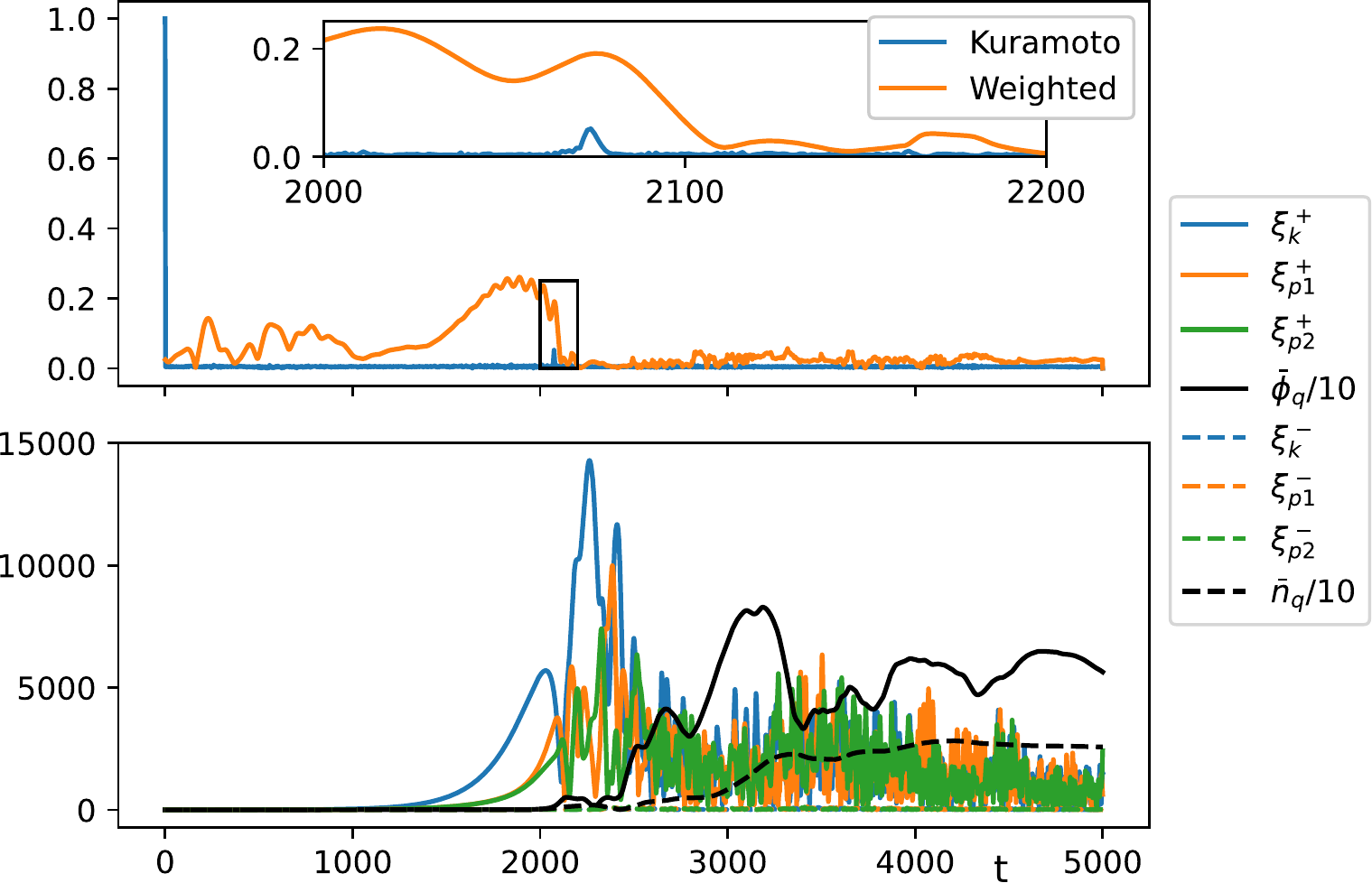}
\par\end{centering}
\caption{\label{fig:order3} The top plot shows the order parameter $r$ defined
in (\ref{eq:z_kuro}) or (\ref{eq:z_weg}) as a function of time for
a DNS. The bottom plot shows the amplitudes of a triad pair with $q=0.5$
and $k_{y}=1.125$ in order to compare with the earlier plots. The
saturation levels for the amplitudes are different because of the
normalization factor $N_{x}^{-1}N_{y}^{-1}$ in front of the nonlinear
term implied in discrete Fourier transforms.}
\end{figure}

\begin{figure}
\begin{centering}
\includegraphics[width=1\columnwidth]{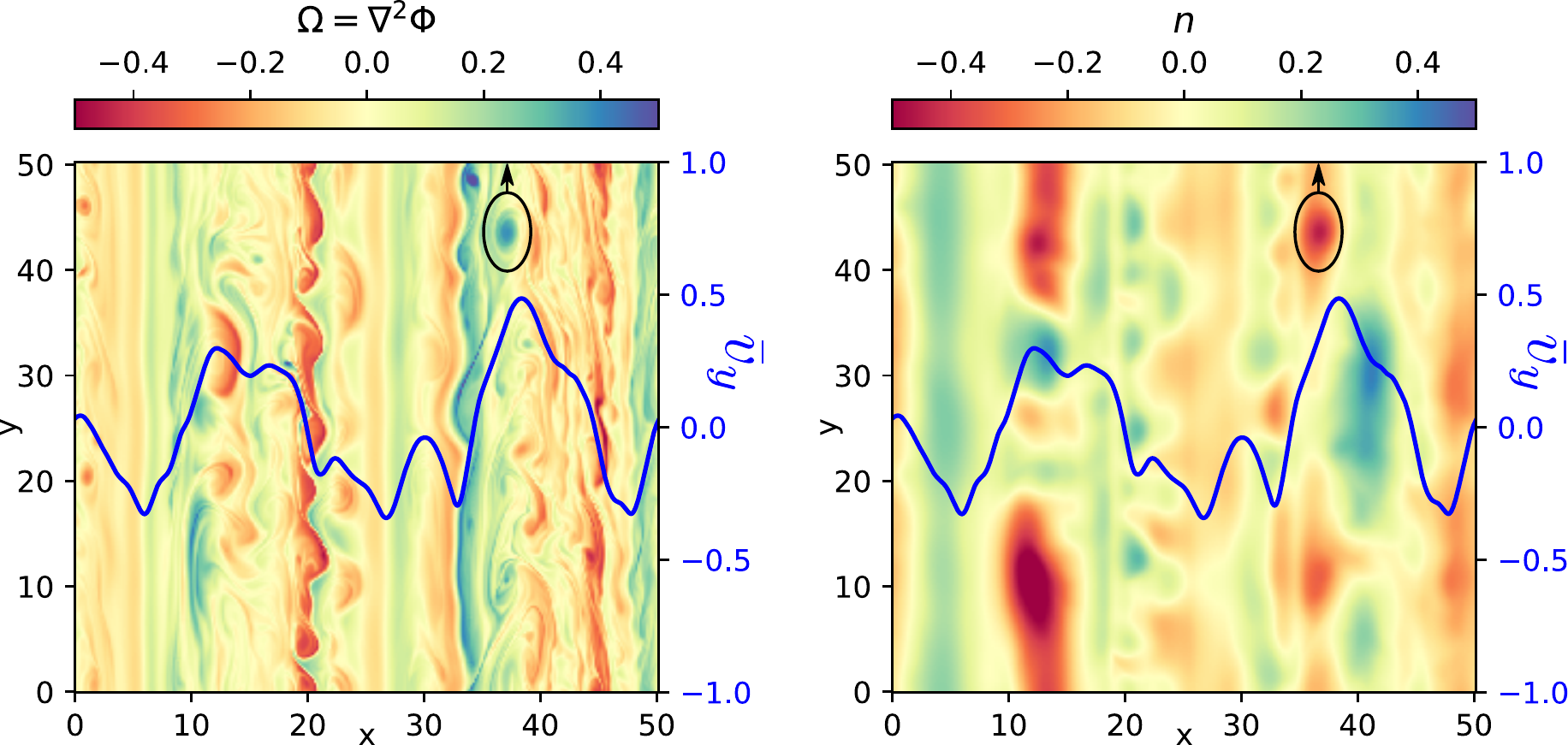}
\par\end{centering}
\caption{\label{fig:snapshots}Snapshots of vorticity and density at t=5000
from DNS. The blue curve in both plots shows the zonal velocity whose
values are given on the right hand axes. An example coherent vortex,
that was moving upwards is encircled.}

\end{figure}

\begin{figure}
\begin{centering}
\includegraphics[width=1\columnwidth]{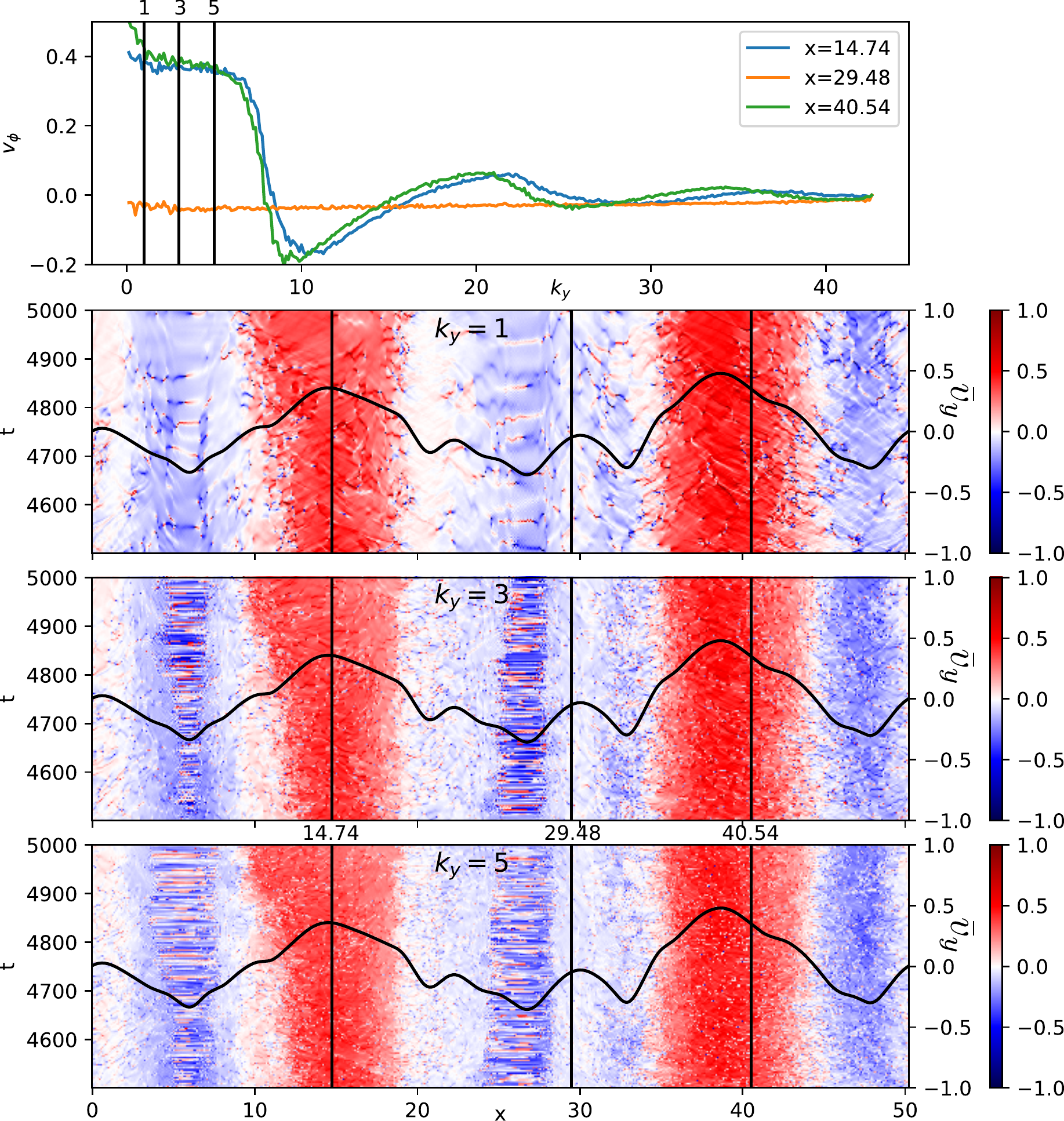}
\par\end{centering}
\caption{\label{fig:phasevel}Profiles of phase velocity as a function of $k_{y}$,
at three different values of $x$ (i.e. $14.74$, $29.48$ and $40.54$)
averaged over $t=[4500,5000]$ shown at the top plot. The three plots
that follow show the detailed time evolution (on the left $y$ axes)
of phase velocity as a function of $x$ for three different values
of $k_{y}$ (i.e. $1$, $3$ and $5$), together with the mean velocity
profile shown for reference (on the right $y$ axes). The phase velocity
is computed using $v_{\phi}=-\partial_{t}\phi_{k_{y}}^{+}\left(x,t\right)/k_{y}$.
The $k_{y}$'s for which the time evolution is given and the $x$'s
for which the phase velocities are shown are marked with horizontal
lines in the corresponding figures. }
\end{figure}

As with all the previous examples of single or multiple triads, or
networks with a particular selection of nodes and triads, we use $C=1$,
$\kappa=0.2$. Since we have a larger range of wave-numbers, we choose
$\nu=D=10^{-4}$, with a box size of $L_{x}=L_{y}=16\pi$ and a padded
resolution of $1024\times1024$. The results show (see figures \ref{fig:order3}
and \ref{fig:snapshots}):
\begin{enumerate}
\item Initial linear growth followed by nonlinear saturation.
\item Formation and finally suppression of nonlinear of convective cells
that transfer vorticity radially.
\item Consequent stratification of vorticity leading to a state dominated
by zonal flows (as in figure \ref{fig:snapshots}).
\item Coherent nonlinear structures (e.g. vortices) that are advected by
the zonal flows in regions of weak zonal shear, get sheared apart
if they fall into a region of strong zonal shear.
\end{enumerate}
Since the wave-like dynamics seems to be primarily in $y$ direction
and reasonably localized in $x$, we can compute the Fourier transform
in $y$ and plot phase of $\xi_{k_{y}}^{\pm}=\chi_{k_{y}}^{\pm}e^{i\phi_{k_{y}}^{\pm}}$
at each $x$, compute $\partial_{t}\phi_{k_{y}}^{\pm}\left(x,t\right)$
in order to compute the phase speeds (see figure \ref{fig:phasevel}).
We can also compute an order parameter as a function of $x$ and $t$
from this data.

While it is clear from \ref{fig:order3} that there is no global synchronization
in direct numerical simulations, the plateau form of the phase velocity
as a function of $k_{y}$ at the radii where it is positive for large
scales, suggest that a process of phase locking similar to soliton
formation in nonlinear Schrödinger equation, where nonlinearity would
balance dispersion is at play for a range of $k_{y}$ values around
the linearly unstable mode. While $\omega/k_{y}$ being the same across
a range of $x$ and $k_{y}$ values is obviously very different from
$\omega$ being the same. However if we note that the nonlinear dispersion
relation takes the form $\omega\left(x,k_{y}\right)=\overline{v}_{\phi}\left(x\right)k_{y}$,
at the lowest order we can see that the frequency in the frame moving
with the zonal flow velocity becomes zero. This is roughly consistent
with what we see in time evolution, where coherent structures like
rotating vortices are advected by zonal flows. In order for such a
detailed structure

\section{Conclusion}

A detailed analysis of triadic interactions formulated in terms natural
frequencies reveals the complex nature of the dynamics of the phases
and amplitudes in the Hasegawa Wakatani system. In particular, it
is observed that a single resonant (or near resonant) triad, including
a pump mode and two other modes, can saturate by adjusting the sums
of phases of its legs ($\psi_{kpq}^{s_{k}s_{p}s_{q}}=\phi_{k}^{s_{k}}+\phi_{p}^{s_{p}}+\phi_{q}^{s_{q}}$)
to be asymptotically constant, resulting in a set of nonlinearly shifted
frequencies and constant amplitudes. When the interactions with zonal
flows are considered, a similar saturation is possible for a single
triad even without the condition of resonance. However this solution
breaks down when we add the triad, which is the reflection of the
original one with respect to the $y$ axis (or the wave-vector $\mathbf{k}$
vector). Instead we observe three different behavior for these triad
pairs as a function of the radial wave number.
\begin{enumerate}
\item For smaller radial wave numbers, we find that the subdominant mode
becomes the dominant one and grows exponentially. We call those unstable
triads. They are associated with unstable subdominant modes.
\item For medium radial wave numbers, after an initial growth phase, the
system saturates with a more or less chaotic evolution, where the
energy goes back and forth between the modes. We call these saturated
triads. They are associated with weakly unstable, or weakly damped
subdominant modes.
\item For large radial wave numbers the system decays to a steady state
solution after a number of limit cycle oscillations. In some cases,
these limit cycle oscillations can continue until the end of the simulation
time. We call these decaying triads (even though they don't decay
to zero but to a constant). They are associated with strongly damped
subdominant modes.
\end{enumerate}
In order to study the dynamics when those triads are connected to
one another, we considered a network formulation where the wave numbers
(or wave number eigenmode combinations) are considered as nodes, and
each triad represents a three body interaction. It is shown while
the zonal flow is almost never dominant in a single triad, when the
whole triad network with a large number of triads is considered, the
zonal modes become dominant \emph{almost in each triad}. Thus, the
system can reach a steady state where the zonal flow dominates as
the other modes decay.

In terms of triadic interactions, as the zonal flow becomes dominant,
it plays the role of a collective mean field, in the sense that for
each mode individual interactions with non-zonal modes start to become
less important compared to the interaction with the zonal flow. This
happens only when the number of triads is large enough so that the
collective wins over the individual. It is interesting to note that
this picture is qualitatively consistent with that of inhomogeneous
wave-kinetic formulation, where the zonal flow is treated as a collective
mean field, and the direct interaction between the modes are either
dropped or modeled with a diffusion operator. This suggests that the
wave-kinetic formulation may hold beyond its range of validity.

Playing with the range of radial wave-numbers of the network model,
we observe that when the range includes only unstable triads {[}i.e.
(i) above{]}, or unstable and saturated triads {[}i.e. (i) and (ii)
above{]} the network system remains unstable. It saturates only when
we include a sufficient range of decaying triads, with subdominant
modes with $\gamma_{p}^{+}<0$. This means that 'local coupling to
damped modes' (i.e. $\gamma_{p}^{-}$ modes even though $\gamma_{p}^{+}>0$)
is not a real mechanism for turbulent saturation. However since the
fact that $\gamma_{p}^{+}<0$ for those modes do not come directly
from dissipation but rather the detailed form of the linear growth/damping
whose form is determined by various parameters including dissipation,
it is correct to argue that in contrast to the Kolmogorov picture
where there is an injection scale, a dissipation scale and the inertial
range in between, plasma turbulence can generate and dissipate energy
in much closer scales, even though one may observe clear power law
scalings.

One of the goals of the current paper was to study the effect of nonlinear
synchronization of drift waves\citep{block:01} on the turbulent cascade
using a framework similar to the Kuramoto model\citep{kuramoto:book:1984},
which has already been attempted using simple models in fusion plasmas\citep{moradi:15,moradi:17}.
We hoped by considering a network of connected triads interacting
with zonal flows we could setup a system that would tend toward synchronization
through slight nonlinear modifications of the frequencies through
their interactions with the zonal flow, playing the role of the control
parameter. However due to particular form of the systematic dependency
of the frequencies to the wave-numbers through the dispersion relation,
such a system does not seem to tend towards synchronization. It should
be checked whether or not the discretization resulting from boundary
conditions, for example in cylindrical geometry change this picture
drastically by impeding resonant interactions\citep{kartashova:94,kartashova_book:10}
especially among large scale modes.

%aipnum4-2.bst 2019-01-14 (MD) hand-edited version of apsrev4-1.bst

\end{document}